%% file: paper.tex
\documentclass[sigplan, screen, nonacm]{acmart}

\usepackage[subtle]{savetrees}

\usepackage{epsfig}
\usepackage{balance}
\usepackage{graphicx} 
\usepackage{subcaption}
\usepackage{tabularx}

\usepackage{xspace}
\usepackage{listings}
\usepackage{verbatim}
\usepackage{booktabs}
\usepackage{colortbl}

\usepackage{nicefrac}
\usepackage{siunitx}

\usepackage{amsmath}
\usepackage{xfrac}

\usepackage{color}
\definecolor{darkred}{rgb}{0.7,0,0}
\definecolor{darkgreen}{rgb}{0,0.5,0}
\hypersetup{colorlinks=true,
        linkcolor=darkred,
        citecolor=darkgreen}

\usepackage{easy-todo}
\usepackage{fancyvrb}
\usepackage{relsize}

\usepackage{algorithm,algorithmicx}
\usepackage[noend]{algpseudocode}

\usepackage[small,compact]{titlesec}
\titlespacing*{\section}{0pt}{4pt}{4pt}%{1*\baselineskip}{1*\baselineskip}
\titlespacing*{\subsection}{0pt}{4pt}{4pt}%{1*\baselineskip}{0.6*\baselineskip}
\titlespacing*{\subsubsection}{0pt}{0pt}{0pt}%{1*\baselineskip}{0.6*\baselineskip}

\input{defs}

\begin{document}

\title{\LARGE SWP: Microsecond Network SLOs Without Priorities}
\author{Kevin Zhao, Prateesh Goyal, Mohammad Alizadeh, and Thomas E. Anderson}
% \author{Paper \#230, 12 pages body, 14 pages total}
\renewcommand{\shortauthors}{X.et al.}

\begin{abstract}
  \input{abstract.tex}

\end{abstract}

\maketitle
\pagestyle{plain}
\input{intro.tex}

\input{background.tex}

\input{limits.tex}
\input{method.tex}
\input{model.tex}
\input{eval.tex}
\input{related.tex}
\input{conclusion.tex}
\def\bibfont{\normalfont}

\bibliographystyle{abbrv}

\bibliography{refs.bib}

\end{document}

%% file: defs.tex
\newcommand{\ie}{{i.e.}, }

\newcommand{\Fig}[1]{Fig.~\ref{fig:#1}\xspace}
\newcommand{\Tab}[1]{Table~\ref{tab:#1}\xspace}
\newcommand{\Sec}[1]{$\S$\ref{s:#1}\xspace}

\newcommand{\Alg}[1]{Alg.~\ref{alg:#1}\xspace}
\newcommand{\ma}[1] {{\textcolor{blue}{MA: #1}}}
\newcommand{\tom}[1] {{\textcolor{red}{tom: #1}}}

\newcommand{\cut}[1]{}

\newcommand*\Let[2]{\State #1 $\gets$ #2}
\algrenewcommand\algorithmicrequire{\textbf{Precondition:}}

\def\compactify{\itemsep=0pt \topsep=0pt \partopsep=0pt \parsep=0pt}
\let\latexusecounter=\usecounter

\definecolor{mygreen}{rgb}{0,0.6,0}
\definecolor{mygray}{rgb}{0.5,0.5,0.5}
\definecolor{mymauve}{rgb}{0.58,0,0.82}
\definecolor{burntorange}{rgb}{0.8, 0.33, 0.0}

\lstdefinelanguage{dhall}
{
  sensitive=true, % keywords are not case-sensitive
  morecomment=[l]{--}, % l is for line comment
  morestring=[b]" % defines that strings are enclosed in double quotes
}

%% file: abstract.tex
The increasing use of cloud computing for latency-sensitive applications has
sparked renewed interest in providing tight bounds on network tail latency.
Achieving this in practice at reasonable network utilization has proved
elusive, due to a combination of highly bursty application demand, faster link
speeds, and heavy-tailed message sizes. While priority scheduling can be used
to reduce tail latency for some traffic, this comes at a cost of much worse
delay behavior for all other traffic on the network. Most operators choose to
run their networks at very low average utilization, despite the added cost, and
yet still suffer poor tail behavior.

This paper takes a different approach. We build a system, swp, to help
operators (and network designers) to understand and control tail latency
without relying on priority scheduling. As network workload changes, swp is
designed to give real-time advice on the network switch configurations
needed to maintain tail latency objectives for each traffic class.
The core of swp is an efficient model for simulating the combined effect of traffic
characteristics, end-to-end congestion control, and switch scheduling
on service-level objectives (SLOs),
along with an optimizer that adjusts switch-level scheduling weights assigned to each class.
Using simulation across a diverse set of workloads with different SLOs,
we show that to meet the same SLOs as swp provides, FIFO would require 65\%
greater link capacity, and 79\% more for scenarios with tight SLOs on
bursty traffic classes.

%% file: intro.tex
\section{Introduction}
The performance of many cloud applications
is becoming increasingly dominated by network tail latency~\cite{qjump,silo,killermicro}.
Even when average case message performance is acceptable,
end-to-end application performance is often
limited by worst case message behavior.
% This effect is becoming more pronounced as applications scale out to more servers
% to meet the ever-escalating demand for cloud services.
On today's data center networks,
the latency of a typical remote procedure call (RPC) or remote memory (RDMA)
operation can vary by several orders of magnitude from average case behavior, even
when networks are kept at low to moderate utilization. Many application programmers
have learned that they simply should not expect
network communication to be fast, except on average~\cite{vahdat}.
% In response, network operators keep average utilization very low,
% advise application developers to avoid correlated behavior at all costs, and
% still frequently miss customer expectations.

Consistently low network tail latency has proved elusive
for a number of reasons. Data center network switches use FIFO queues
where interactions between traffic types can dominate tail behavior.
Data center network traffic is highly bursty, meaning
that average and tail load can diverge radically, even within a single traffic class.
While link speeds are rapidly scaling up,
with 100\,Gpbs and even 400\,Gbps links becoming standard~\cite{broadcom,switchtrend},
% Extra capacity should make it easier to achieve low latency, but 
application demand is scaling up even more rapidly~\cite{jupiter}. Faster links means that a
substantial fraction of data center network traffic completes within a few round trips,
reducing the portion of traffic subject to congestion control~\cite{aeolus}
and further complicating efforts to keep queues small.

While priority scheduling and admission control can achieve guaranteed low latency~\cite{qjump,silo},
that only works for a small slice of well-behaved network traffic and it comes at the cost of much worse
variability for the remaining traffic. Our interest is in providing tight bounds
for all traffic classes---in particular, we show that priority
scheduling is needlessly aggressive in many situations.
Likewise, endpoint congestion control attempts to keep queues bounded
and small to achieve better response times for very short transfers,
but this often comes at a cost of worse latency variation for medium-sized transfers and a loss
of throughput for long flows~\cite{dctcp,hpcc,homa}.
Our interest is in supporting tight latency bounds
across traffic classes and mixtures of short and medium-length flows.

% \ma{is bucket the same as class? why use two words?}
We assume that the network operator splits traffic into a small number of
classes, where each traffic class has a service level objective
(SLO)~\cite{mogulnines}.
We assume these SLOs are probabilistic rather than absolute, matching the probabilistic
SLOs for network and service availability that operators already provide. At datacenter
scale with hundreds of thousands of servers, deterministic SLOs are impossible
except for a small fraction of datacenter traffic~\cite{qjump,silo}. The SLOs we
consider in this paper provide tight bounds, such that 99\% of messages (or flowlets~\cite{flowlet}), regardless of size, are delivered within a small integer factor of
the time they would take on an unloaded network.
% For a small message, that means delivery within 1.5 $\times$ the round trip 
% time, with proportionately more slack for longer transfers. \pg{I think this is assuming the best FCT is 0.5RTT, but if I remember correctly Kevin measures FCT by looking at the first packet sent and the last ACK received (same as HPCC).} 
We note that our tools are generalizable to weaker or stricter SLOs,
and to different SLOs for different transfer sizes.

Further, we assume the network operator continuously gathers the distribution of
message lengths and interarrival times (burstiness) of messages within each traffic class.
This can be accomplished through traffic sampling at the RPC, socket, virtual machine, or RDMA level,
% \ma{sampling at the RPC level might be most useful since RPCs correspond precisely to the messages we care about}
with periodic updates of traffic estimates.
We assume no change to the standard socket level API---the network interface or switch
does not have access to message/flowlet length (except in retrospect). Thus, we do not consider
switch scheduling mechanisms that use flow length to favor short messages,
such as shortest remaining flow first~\cite{homa},
or to implement deadline first scheduling~\cite{pifo,spfifo}.
We also assume widely used endpoint congestion control mechanisms, specifically
DCTCP~\cite{dctcp} and HPCC~\cite{hpcc}.

At the switch level, we assume only standard configurability of modern datacenter
switches such as Broadcom's Tomahawk 4~\cite{broadcom} or Intel's Tofino 2---the operator
can assign strict priorities or scheduling weights to each of a small number of traffic classes.
A class with a normalized weight of 0.4 and with a queue of packets,
for example, will have its packets scheduled at least 40\% of the time.
% \pg{Minor point, I think the weight convention in Tofino2 is different, traffic class have integer weights ($w_i$ for class i) and are not normalized. According to our definition of weights, weight of class i = $\frac{w_i}{\sum{w_i}}$. We are (maybe incorrectly) assuming setting any normalized weight is possible.}
Within each class, we assume FIFO scheduling. We also explore the potential benefit of switch
programmability by considering the use of a calendar queue~\cite{cal-queues} to
implement fair queueing within each traffic class.
We make the simplifying assumption of considering a single bottleneck at a time,
leaving multiple bottlenecks for future work.

With these knobs, we build swp (SLOs without priorities) to efficiently determine
if target network SLOs can be met given estimated load, burstiness, and flow length distribution
for each traffic class. If SLOs can be met, swp provides switch configuration weights
for each traffic class. swp can also be used prospectively, to
evaluate the feasibility of target SLOs given potential future traffic changes, e.g.,
due to the rollout of a new application, an anticipated spike in traffic, or
a prospective change in endpoint congestion control policy.

swp provides two benefits. First, instead of giving priority to whichever traffic class has
the tightest deadlines, we allow the scheduling weight for each class to be the \emph{minimum}
necessary to meet its SLO given its burstiness, message size distribution, and utilization.
% fixed by changing set to allow
% \pg{I don't think that swp finds the minimum necessary weights, my understanding is that swp just finds whether a weight configuration (sum of weights = 1) exists which will satisfy SLOs. The minimum weight for a class to meet SLO doesn't really depend on about whether the SLOs for other classes are being met are not.} \ma{Yes this doesn't seem accurate}\tom{I changed the above to allow, which I think maybe works?} \ma{i don't think you need 'normalized'; it's a detail.} \ma{i think strictly speaking, swp may not find the minimum necessary weights, but `we allow' is better. the point being made here in contrast to priority scheduling is important} 
The more bursty a traffic class, and the tighter its SLOs,
the greater headroom is needed above and beyond its average utilization, in order to meet its SLOs.
In many circumstances, a set of scheduling weights can meet the SLOs of each class,
where strict priorities, or endpoint congestion control alone,
would not be able to.  If all classes can meet their deadlines,
excess capacity is distributed to (approximately) minimize the chance of an SLO violation. % \pg{I think extra weight is probably not defined well here. doesnt the swp optimization always work with sum of weights = 1?}

Second, if the bursts of traffic in one class are uncorrelated
with the traffic in other classes, we can overcommit weights relative to what each class would
need if it was running in isolation.
Since each class does not require its entire headroom all the time, a link multiplexed between multiple traffic classes can statistically support a higher load than would be possible otherwise.
% Each class will need some headroom to meet its SLOs, which
% is used only when necessary.  From a statistical perspective, that headroom improves the margin
% of error for all other classes, particularly those with somewhat less strict SLOs, allowing
% them to support a higher load than they would be able to otherwise. \ma{A little confusing. consider something like: Since each class does not require its entire headroom all the time, a link multiplexed between multiple traffic classes can statistically support a higher load than would be possible otherwise.}
One can think of this as the equivalent of the use of slack in deadline scheduling,
but computed on traffic aggregates. This extra capacity is not completely free: if
a particular traffic class exceeds its expected utilization or burstiness, it can cause missed
deadlines for other traffic classes.
% if true - We present our results both with and without this optimization.

% \ma{It seems that (1) and (2) are benefits of swp, and (3) is the main challenge we solve.}

A key barrier to swp is the efficient computation of tail latency behavior
% \ma{tail latency or message completion time would be better. it's more than just computing tail queue behavior}
given a particular switch configuration and message arrival pattern. The state of the art
would be to simulate (or directly observe) the queueing behavior in detail for a sufficiently long
sample to gain statistical reliability, repeated
for each possible switch configuration.  The inner loop of that calculation is gated by the operational
behavior of the congestion control mechanism---the queue length at each instant in time,
what packets are marked (in DCTCP~\cite{dctcp}) or congestion information returned (in HPCC~\cite{hpcc}),
when that information would reach the endpoint, how the endpoint would react, etc.
% for every observed state of the queue. \ma{cut `for every observed state of queue'; not wrong but just to simplify}

Instead, for our setting, we only need sufficient accuracy to predict tail latency
SLOs. We create a high-level abstract model of each endpoint congestion control algorithm,
where the control loop operates at a time lag but with perfect information
about the remote queue. This simplified model speeds up execution time by 50-80$\times$
relative to ns3. We calibrate the models (one each for HPCC and DCTCP) using ns3 simulations,
and we show that the resulting models are accurate enough to provide a basis for
computing switch configurations to meet probabilistic tail latency SLOs.

We evaluate the robustness of swp in simulation.  We sample randomly among
plausible scenarios of three and five traffic classes, with varying
utilization, burstiness, traffic size distribution, and SLO tightness. We use
swp to determine the optimal configuration to meet the SLO
for each scenario with the least aggregate bandwidth.
We then repeat the same scenario assuming a single FIFO queue, multiple FIFO queues (one
per traffic class) with weights assigned by swp, and an idealized
% fair queue per traffic class with weights assigned by swp 
hierarchical fair queuing scheduler with per traffic class weights assigned by swp.

% \ma{using the word configuration to refer to both the network config. and traffic scenario is very confusing. i replaced the latter with `scenario'}

Averaged across all five-class scenarios, FIFO requires 65\% more link capacity
to accomplish the same SLOs as swp. This benefit increases to 79\% for more
challenging scenarios where at least one traffic class has a tighter SLO and is
relatively bursty. Using swp with fair queueing gains another factor of two in
link capacity on average across all scenarios, while still meeting SLOs.

%% file: background.tex
\section{Background and Motivation} \label{s:background}
In this section, we consider the limitations in the use of priorities
and traffic shaping to achieving tight service level objectives for multiple
classes of traffic.

We begin by defining terms. Our goal is to provide tight, probabilistic bounds on tail latency
for latency-sensitive traffic in data center networks. To distinguish
connections that may be reused, we consider each message separately,
e.g., each remote procedure call (RPC), remote memory operation (RDMA), or independent
data transfer, where latency is measured as the time to complete the transfer. 
We define message latency as the time to complete a message transfer,
including transmission, propagation, and queueing delay, from when the first packet is
available to be sent until the last packet arrives at the destination. 
% \pg{check.} \ma{added transmission time}
In particular, we include in the latency any queueing at the end host queue
needed for traffic shaping or congestion control.

We define message slowdown as the message latency divided by the minimum
latency on an unloaded network. For example, in a network with a round trip
propagation delay of 10$\mu$sec and 100Gbps links, the minimum latency
for a 125KB transfer would be 20$\mu$sec. 
We can also define tail slowdown behavior separately 
for different message sizes to prevent swp from optimizing 
for small messages at the expense of medium-sized or long messages. 

The specification of the tail probability bound on message slowdown 
is configurable in swp, but our aim is to provide bounds that
are tight enough for application developers to largely ignore tail effects. 
Thus, we focus in this paper at tail message slowdowns, across different message sizes, 
of a small integer multiple of the best case behavior.

It is impossible to provide bounds on the slowdown for any shared resource
without some characterization or bound on the arrival process of requests.
Mogul and Wilkes call this Customer Behavior Expectations (CBE)~\cite{mogulnines}.
We assume only a probabilistic characterization, provided by an ongoing measurement
of application network usage. Some prior attempts at providing network quality of service,
such as IntServ~\cite{rsvp}, assume users provide hard limits on their traffic 
demands which can be guaranteed (or denied) at runtime depending on current traffic
conditions. For many cloud applications, however, network traffic demand 
is a dynamic property at varying time scales, resistant to deterministic limits. 
At any point a flash crowd may appear, and the system should be configured to handle
these within its promised probabilistic performance envelope.
% bounds, provided the bursts happen infrequently. \pg{do you mean to say there should be some bound on burst frequency. I think infrequently is not well defined here.} \ma{fwiw I am ok with the paragraph. The goal isn't to make a formal statement; it's to give intuition. We could even get rid of the `provided the bursts happen infrequently' part. The likelihood of bursts and their frequency is taken into account in the probabilistic bounds; i.e., it's not a separate requirement, it was already considered when committing to the probabilistic performance bounds. All that is required is that the traffic conforms to the statistical characterization that was used to derive the bounds.}

We assume traffic is inherently bursty,
with traffic measurement conducted on a long enough interval to allow us to construct
a model of the traffic behavior. For each traffic class, we assume a sampled measurement 
process of the distribution of message sizes and message 
interarrival distribution to characterize traffic from that class. 
% \ma{avg. intensity is redundant; it's determined by distribution of message sizes and interarrival times}

Following the terminology in Mogul and Wilkes~\cite{mogulnines},
a Service Level Objective (SLO) lets a provider describe in precise terms the quality of
service it aims to give its users. By writing an SLO, an operator 
codifies the properties that can be relied on,
guarding each party against potentially mismatched expectations.
A building block for SLOs is the service level indicator (SLI), specifying
some metric of interest, such as the tail latency for 
small requests or average throughput for larger requests. For a particular
class of traffic, the SLO specifies a bound for the relevant SLI and can
combine bounds on different SLIs in a conjunction.  For example, we
can specify that
all memcache traffic, regardless of message size, has a tail slowdown of no more than three,
at least 99\% of the time. swp provides a small specification language for the network
operator to specify its SLIs and SLOs. swp determines
whether the SLO can be met given competing classes of traffic and network capacity.

%% file: limits.tex
\input{fig1}

\subsection{Limits of Priority Scheduling}
To motivate the approach taken by swp, we develop a simple experiment
to characterize the limitations of priority scheduling for providing feasible SLOs
for non-priority traffic. Our evaluation of swp explores the parameter space more fully.

We start with a large number of servers connected by 100 Gbps links and a round-trip delay of 10 \si{\micro\second}.
We focus on a single bottleneck, located near the destination, with traffic split
between foreground (higher
priority) and background (lower priority) traffic. Message sizes for both foreground
and background traffic are drawn from the Homa W3 distribution~\cite{homa}, taken as a sample
of all messages in a Google data center.
% To reflect that operators
% often restrict the maximum size of high priority messages because of the impact of high priority
% traffic on lower priority traffic, we restrict the foreground
% class to messages that fit within the bandwidth-delay product of the network, that is,
% no larger than 150KB.  
% Background messages are drawn from the entire Homa W3 distribution.
We assume the two traffic classes are independent of each other, but within each traffic
class, flows have log-normal interarrival time distributions with a shape
parameter $\sigma$ of two. Further, we assume HPCC congestion control with an initial window size of the
bandwidth-delay product~\cite{hpcc}.
% this means that foreground traffic is sent at
% line rate while background traffic is controlled after the initial window.

For this experiment, we focus on 99\% tail message slowdown for messages
less than the bandwidth-delay product. In this workload, these
account for the large majority of requests but less than half of the total bytes transferred.
To test how well each configuration insulates foreground traffic
from background traffic, we fix the foreground
utilization at 10\% and vary the background utilization from 20\% to 80\%.
We choose a target slowdown of 2.5$\times$ for the SLO, that is, a tail
latency for
% small messages of 60$\mu$sec, and 
125 KB messages of 50 \si{\micro\second}.

\if 0
    There is a single bottleneck with an infinite queue, and we assume there are
    two traffic classes---one at high priority, and the other at low priority.
    For brevity we refer to these classes as ``priority'' and ``background,''
    respectively.
    Further, we assume an end-to-end congestion control algorithm that can
    distinguish between the priority traffic and the background traffic: priority
    traffic only reacts to congestion due to itself, while background traffic
    accounts for congestion due to either traffic class.
    This assumption is optimistic, but it will not affect our conclusions about
    priority queueing.
    As for the workload, both classes have a flow size distribution similar to
    workload W3 from Homa~\todo{cite}, and the flows have log-normal interarrival
    time distributions with shape parameter $\sigma$ equal to two.

    Our metric of interest is the P99 flow completion time slowdown of small,
    latency-sensitive flows, which we will say are flows smaller than 1024 bytes
    (in our workload, these flows account for more than half the number of total
    flows).
    %
    We use this metric to set an SLO for the priority traffic, stipulating that the
    tail slowdown of priority flows may not exceed 2.5$\times$.
\fi

%

\Fig{prio} compares the 99\% tail message slowdowns for four configurations.
First, we pair endpoint congestion control with a FIFO queue at the switch, shared between
both foreground and background traffic (\Fig{prio-baseline}).
As expected, using a shared queue means that background traffic can interfere
with foreground tail latency. At low load, both foreground and background
traffic can achieve the target SLO. As background load increases, endpoint
congestion control limits the effect of the background traffic on the
foreground traffic, but at high enough loads, the unconstrained portion of the
traffic (within the initial window) can impact foreground tail delay.

Note that the tail latency for foreground and background traffic differ.  This is because
the inter-arrival distribution is heavy-tailed and hence bursty \emph{within its own traffic class},
but uncorrelated across traffic classes.
% each is heavy-tailed but only correlated within its own traffic class.
% \ma{To be precise, seems useful to emphasize that the issue here is that the {\em inter-arrival} distribution is heavy-tailed and hence bursty within each traffic class. (I think it matters less that the message sizes are heavy-tailed for this observation; if the traffic was Poisson arrivals with heavy-tailed flow sizes, the foreground and background would have looked the same.}
Conditional on a small background message arriving at the switch, it is more likely
that other background flows will already be queued at the switch; this is the definition
of heavy-tailed behavior. Foreground traffic is more likely to encounter other foreground flows;
however, the foreground load is lower so that occurs less often.
% The foreground traffic mix also excludes long flows which by definition have a larger impact on tail queue length than short flows.

% and we see that end-to-end congestion control
% alone cannot preserve a tight SLO under increasing interference. \ma{I'm assuming the tail slowdown is different for priority and background because the CC is priority-aware. This is an unsual assumption for flows sharing a FIFO queue, so we may want to add a footnote explaining this point in relation to Fig. 2(a).}
%
% In modern networks, priorities are often used to provide low latency SLOs for a
% subset of traffic.
%
Next, we add strict priority scheduling at the switch, where foreground traffic
always takes precedence over background traffic (\Fig{prio-prio}).
This achieves the SLO for the foreground traffic regardless of the
background traffic intensity, but only at the cost of much higher
small message response time for background traffic. Note that the y-axis is rescaled to show
the effect. Because the foreground traffic takes priority, if the background
traffic arrives during a burst of foreground traffic, it will experience
head-of-line blocking---no progress until the burst is cleared.  This has
a measurable effect on background tail latency, even though the foreground
traffic uses only 10\% of the link in aggregate.
% \ma{could cut last sentence to save space}

\input{fig2}

Although not shown because of the y-axis rescaling, the foreground traffic
achieves its SLO with considerable room to spare; in this scenario, prioritization
is overly conservative, needlessly harming background tail latency.
% \ma{also probably cut for space; the point has been made already}

\if 0
    As a consequence, the high priority traffic is effectively a head-of-line
    blocker for all other traffic whenever it has packets queued at the switch,
    with no regard for the specific requirements of its SLO.

    When we equip the switch with strict priority queueing, the priority traffic
    can meet its SLO regardless of the amount of background interference
    However, this performance isolation is won at a large cost to the background
    traffic, whose head-of-line blocker utilizes the link 10\% of the time, causing
    the background traffic's P99 slowdown to skyrocket.
\fi

The third graph in the figure (\Fig{prio-classes}) considers the impact of traffic class weights
on SLOs. In this case, the switch has separate FIFO queues for each traffic class, and schedules
among each queue according to its weight if both queues are occupied.  When only
one queue is occupied, that queue is scheduled.
To our knowledge, there are no well-established guidelines for how
to set traffic weights. Instead, data center operators act by trial and
error---adjust weights to meet a specific SLO in a specific situation.

We model this by setting the foreground scheduling weight so that it meets its SLO
tail latency target even for the highest level of background traffic (80\% load),
with a small margin of error.
% (Note that with 80\% average load, there will be periods of no load; with20\% load, there will be periods where the queue is busy for multiple round trips. \ma{didn't follow why we talk about 20\% load here and multiple round-trips})
A single weight is able to insulate the foreground tail latency across the entire
range of background traffic
intensity, unlike CC+FIFO. Likewise, although the background traffic is unable to
meet the target SLO, it experiences much better tail latency than with priority scheduling.

We could do even better for reducing background tail latency if we were to know
in advance the average traffic intensity of the background traffic.  At low
background utilization, the traffic weights chosen above are needlessly strict.
With less competing background traffic, we can afford to give the foreground traffic
less weight---less headroom above its traffic demand---and still meet its SLOs. This is
because most of the time the foreground traffic will arrive at the switch to find
the background queue empty, improving its overall performance. This insight---that we
should set weights given knowledge of the foreground and background traffic characteristics---lies
at the core of the design of swp.

Finally, we consider the case of a programmable switch capable of implementing
fair queueing~\cite{fq} through the use of calendar queues~\cite{cal-queues}.
We assume a hierarchical setting, where weights are used to choose among traffic classes
when both have traffic present.  Within each traffic class, separate calendar
queues are used to implement fair queueing among flows with traffic queued within
that class. Fair queueing allows better isolation between competing flows within
the same traffic class, by eliminating head-of-line blocking. With a diversity
of message sizes, with FIFO queueing a short message can be delayed behind packets
of a longer flow.  A fair queued system allows the short message to be scheduled
earlier.
Recall that we give the foreground traffic just enough weight to meet the SLO
in the presence of background traffic at 80\% load, with the remaining capacity
given to the background traffic.
With fair queueing, the foreground traffic requires so little weight to meet
the SLO that the background traffic even outperforms it (\Fig{prio-ideal}).
In this case the SLO could be further tightened without incurring additional
SLO violations.
In \Sec{method-optim}, we describe a more sophisticated and more general
optimizer that can find weights to meet distinct SLOs for multiple traffic
classes.

\subsection{Deterministic Latency Bounds} \label{s:deterministic}
Prior work in bounding network tail latency has used endpoint traffic shaping
and worst case analysis to derive deterministic guarantees. These guarantees
are, in a sense, stronger than what is provided by swp, in that they provide bounds
on worst case behavior for queueing even at the tail, as long as packets are not corrupted in flight. On the other hand, particularly for large scale systems with
bursty traffic, the bounds are substantially looser than what swp can provide.

For example, QJump~\cite{qjump} and Silo~\cite{silo} apply a send-side leaky bucket
filter to constrain the worst case load in the network.  Even if all nodes send a burst
at exactly the same time, the leaky bucket will constrain the worst case queueing
at the bottleneck to, roughly, the bucket size times the number of senders, provided
that the switch is configured to give priority to these latency-sensitive packets.
This follows from a classic result due to Parekh and Gallager: if 1) admission to a
network is governed by leaky bucket, 2) the network uses fair queueing, and 3) the
arrival rates are constrained to ensure stability, then the delay in the network
can be bounded~\cite{parekh-gallager}.

There are two important limitations that led us to use a probabilistic,
rather than a deterministic, model. First, most datacenter network traffic is
highly bursty~\cite{dc-traffic}. When source bursts can exceed the bucket size,
a leaky bucket mechanism will impose an additional queueing delay at the source,
to prevent bursts from one node from compromising the SLO's provided to another.
Second, the latency bound scales in proportion to the product of the
bucket size and the number of senders. As the number of possible senders increases,
the allowable burst must decrease proportionately to keep the SLO constant.
For example, for a small data center with 1000 servers connected by 100Gbps links
and a bucket size of 10KB buffer per server, worst case latency is nearly a millisecond.

\if 0
    With leaky bucket, application bandwidth demands are modeled as a baseline (the
    leak) and a maximum burst size (the bucket).
    In effect, it performs admission control at the sources, and allows the
    separation of total delay into two components: the delay at the source, and the
    delay in the network.
    This result has previously been used by QJump~\todo{SILO?}, which combines
    leaky buckets with switch priorities to bound the network latency of
    high-priority traffic~\cite{qjump}.

    However, unaddressed in Parekh and Gallager's analysis is the other delay
    component---the delay at the source imposed by the leaky bucket.
    First, assume a model of leaky bucket in which nonconforming packets are FIFO
    queued at the host rather than dropped and retransmitted.
    If we further assume a Poisson arrival process and exponentially distributed
    packet sizes \ma{packet or flow sizes}, host-side delay can be estimated using standard queueing analyses. But what if the arrival process is not a Poisson process?
    %
    In particular, datacenter traffic has previously been shown to have
    approximately log-normal interarrival time distributions~\cite{dc-traffic},
    which yields a much burstier traffic pattern, and which makes standard
    analytical techniques much more difficult to apply.
\fi

This places the network designer in a bind. Tighten the bucket size, and more
delay is experienced at the source; loosen the bucket size, and the worst case
network queueing goes up. We illustrate this with a simple experiment
(\Fig{leaky}).  First, we consider the send side queue. In \Fig{leaky-cdf}, we
assume exponentially distributed message sizes, with an arrival rate 5\%
smaller than the leaky bucket rate $r$. We set the bucket size to yield
reasonable send side tail latency with Poisson arrivals, and then consider what
happens when we shift to moderately bursty arrivals (log-normal
with a shape parameter of 1.5). Tail delays with the more realistic
log-normal distribution are an
\emph{order of magnitude} higher than would be predicted under the analytically
tractable model.

\if 0
    To quantify the effect of a burstier traffic pattern on host-side delays, we
    first implement leaky bucket in simulation and validate its predictions under
    Poisson arrivals against a simple average delay analysis due to Bertsekas and
    Gallager~\cite{data-networks}.
    Then, we change the interarrival time distribution to log-normal with a shape
    parameter $\sigma$ equal to 1.5~\todo{justify}, and we observe the effect on
    tail host delay.
    The result is depicted in \Fig{leaky-cdf}.
    In this experiment, we set the arrival rate to be 5\% smaller than the leaky
    bucket rate $r$, and we set a bucket size $b$ that yields reasonable tail
    delays for exponential interarrival times.
    We find that tail delays with the more realistic log-normal distribution are an
    \emph{order of magnitude} higher than would be predicted under the analytically
    tractable model.
\fi

To compensate for burstier traffic, we can increase the bucket size, as shown in \Fig{leaky-bsize}---source queueing delay decreases. Unfortunately, this only
trades queueing at the host for queueing further downstream.
\Fig{leaky-qlen} shows the distributions of downstream queue lengths at
different bucket sizes. For each distribution, the median, the quartiles,
and a rotated kernel density estimation is drawn.
We see that as the bucket size increases, so too does the tail of downstream
queueing.
% \ma{This is very good but what would really nail it is if we could show that even with fine-tuning the bucket size, we can't get a good outcome in a scenario like the one in Fig. 1. I'm thinking of a part (d) figure that is similar to the ones in Fig. 1 but uses the  smallest bucket size that allows the foreground to meet its SLO at 80\% load. Not a high priority -- just a nice to have if we have time}

\if 0
    %
    To see this, we attach a second FIFO queue to the output of the leaky bucket
    traffic shaper, also serviced at rate $r$, and we measure the length of that
    queue each time a new packet arrives---this value directly corresponds to the
    delay the incoming packet will experience.
    %

    %

    %
    With 40 Gbps links, 1000 nodes \ma{What does this mean? There was no notion of `node' in the setup in this section.}, and a bucket size of 150 KB, delay bounds under
    Parekh and Gallager's model are in the tens of milliseconds\,---\,far too large to
    be of practical use. \ma{It might make sense to add the end-to-end tail latency to Fig 1(b), showing that as the host delay goes down, the end-to-end delay increases and there's a sweet spot for bucket size but the p99 delay is still pretty bad even at the best possible bucket size. Fig 1(c) is basically the explanation of this result.}
\fi

%% file: fig1.tex
\begin{figure*}[t]
    \centering
    \begin{subfigure}[t]{0.24\linewidth}
        \centering
        \includegraphics[width=\textwidth]{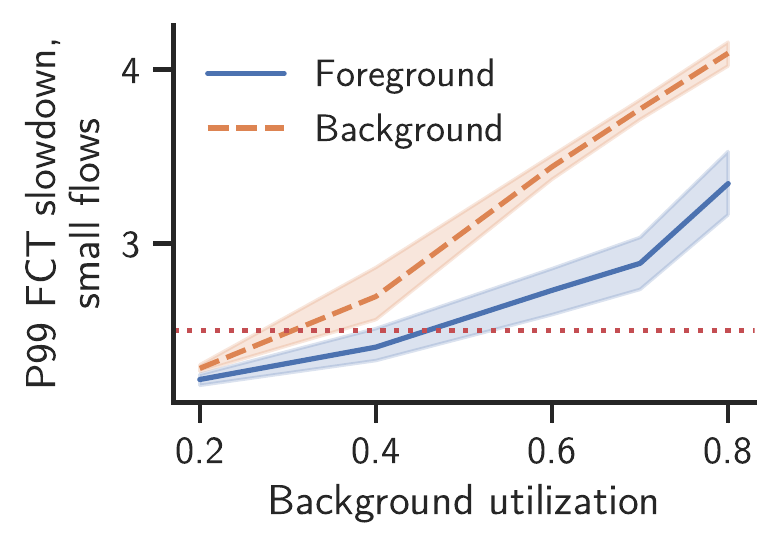}
        \caption{\bf CC + FIFO}
        \label{fig:prio-baseline}
    \end{subfigure}
    \hfill
    \begin{subfigure}[t]{0.24\linewidth}
        \centering
        \includegraphics[width=\textwidth]{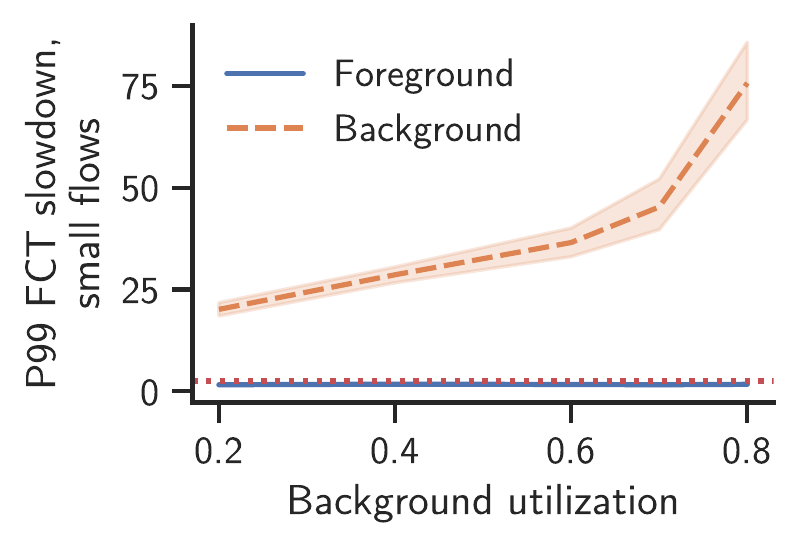}
        \caption{\bf CC + PQ}
        \label{fig:prio-prio}
    \end{subfigure}
    \hfill
    \begin{subfigure}[t]{0.24\linewidth}
        \centering
        \includegraphics[width=\textwidth]{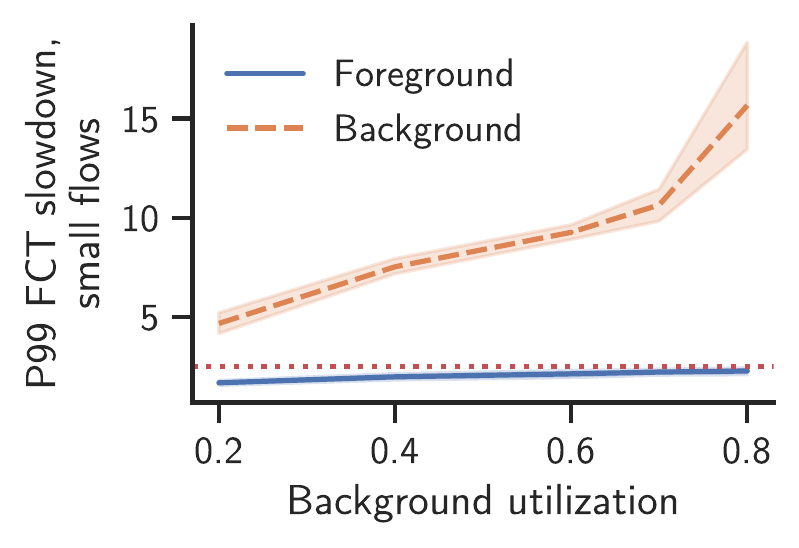}
        \caption{\bf CC + TC}
        \label{fig:prio-classes}
    \end{subfigure}
    \begin{subfigure}[t]{0.24\linewidth}
        \centering
        \includegraphics[width=\textwidth]{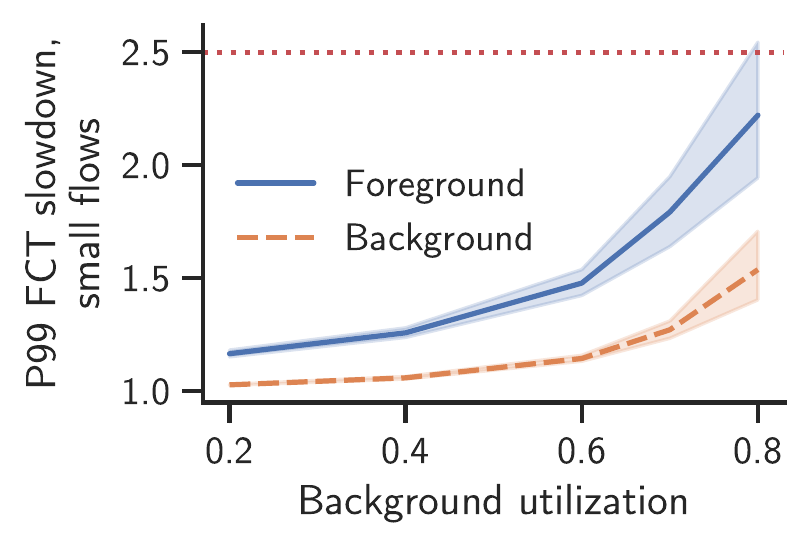}
        \caption{\bf FQ + TC}
        \label{fig:prio-ideal}
    \end{subfigure}
    \vspace{-2mm}
    \caption{\small {\bf Tail (99\%) message slowdown for the Homa W3 workload~\cite{homa}.
            Foreground traffic is set to 10\% of the link bandwidth,
            with variable background traffic.
            The dotted horizontal line shows a 2.5$\boldsymbol{\times}$ SLO, and
            the y-axis scale varies between graphs.
            Shaded regions represent the 95\% confidence interval. We compare
            endpoint congestion control (CC) and a FIFO queue at the switch; CC and
            strict priority queueing (PQ); CC and per-traffic class (TC) scheduling weights at the switch, tuned to meet the foreground SLO with 80\% background traffic;
            scheduling weights and per-traffic class fair queueing among queued flows.
            While priority scheduling can provide latency guarantees for foreground traffic,
            it leaves the tail latency of background traffic higher than necessary.
        }}
    \label{fig:prio}
    \vspace{-3mm}
\end{figure*}

%% file: fig2.tex
\begin{figure*}[t]
    \centering
    \begin{subfigure}[t]{0.25\linewidth}
        \centering
        \includegraphics[width=\textwidth]{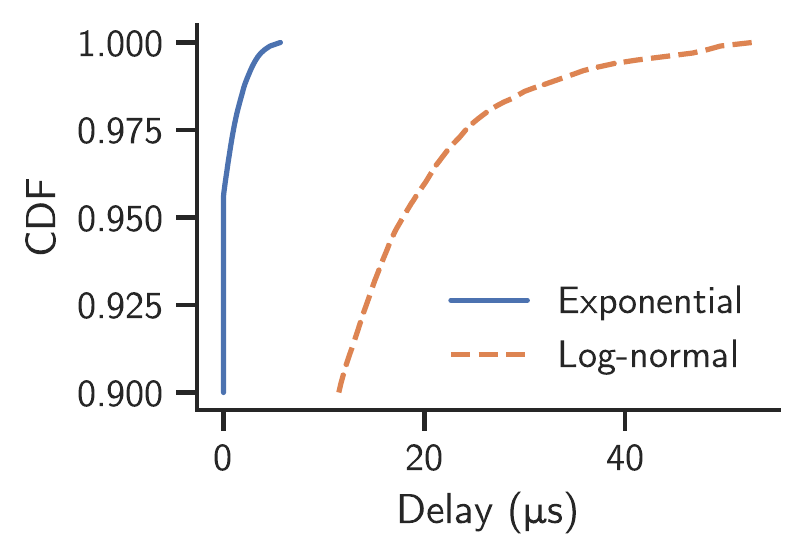}
        \caption{\bf Sensitivity to burstiness}
        \label{fig:leaky-cdf}
    \end{subfigure}
    \hfill
    \begin{subfigure}[t]{0.24\linewidth}
        \centering
        \includegraphics[width=\textwidth]{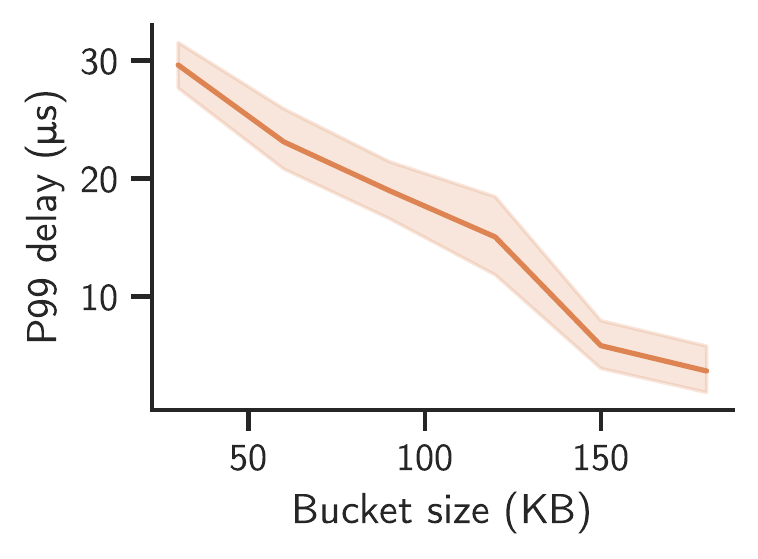}
        \caption{\bf Adjusting the bucket size}
        \label{fig:leaky-bsize}
    \end{subfigure}
    \hfill
    \begin{subfigure}[t]{0.26\linewidth}
        \centering
        \includegraphics[width=\textwidth]{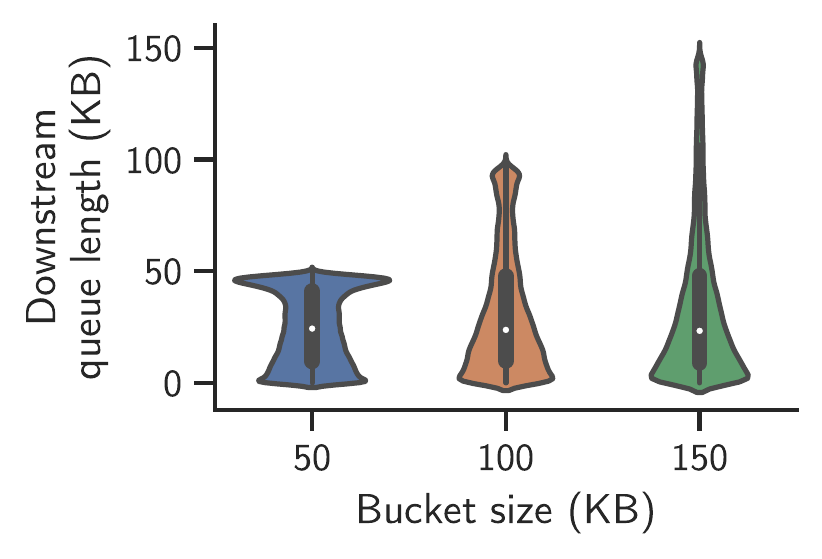}
        \caption{\bf Effect on downstream queues}
        \label{fig:leaky-qlen}
    \end{subfigure}
    \vspace{-2mm}
    \caption{\small \bf
        When a workload is bursty, leaky bucket parameters must be tuned to
        preserve low tail delays at the host. However, this will in turn
        increase the tail of downstream queue lengths (as seen by an arriving
        packet).
    }
    \label{fig:leaky}
    \vspace{-3mm}
\end{figure*}

%% file: method.tex
\section{The swp Methodology}

\begin{figure}[t]
    \centering
    \includegraphics[width=0.8\linewidth]{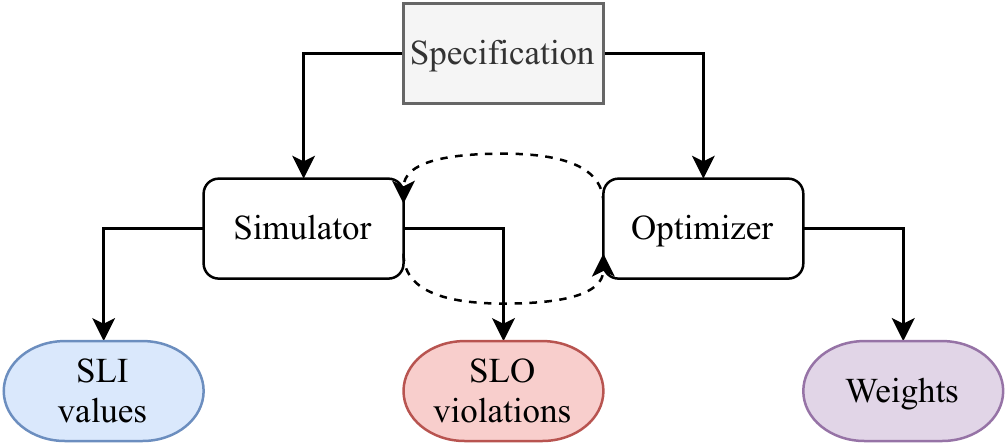}
    \caption{\small \bf
        The swp workflow. Rectangular boxes indicate user inputs, rounded boxes
        indicate swp components, and ovals indicate outputs. Dashed lines show
        the optimization loop for finding weight allocations.
    }
    \label{fig:swp-workflow}
    \vspace{-2mm}
\end{figure}

Our goal with swp is to build a tool to aid network operators with configuring
their network switches to achieve tail latency SLOs for multiple classes of traffic.
Specifically, swp

\begin{enumerate}
    \item allows users to specify a network configuration, a set of traffic
          classes, and their SLOs,
    \item automatically finds switch weights for meeting those SLOs (if possible), and
    \item provides descriptive answers to what-if style questions about SLO
          behavior.
\end{enumerate}

We begin with a brief overview of the swp workflow (\Sec{method-overview}).
Then, we show how to parameterize the network model and specify a traffic
class's SLOs (\Sec{method-spec}).
Lastly, we discuss how simulation outputs are used as input to a downstream
optimizer, and we describe the one we use for automatically finding switch
weights (\Sec{method-optim}). The network simulator is described in detail in \Sec{model}.

\subsection{Overview} \label{s:method-overview}

\Fig{swp-workflow} depicts swp's workflow.
For each traffic class, the user provides information about its flow sizes and
interarrival times. These can be in the form of a trace, a measured cumulative
probability distribution, or a generator function.
Once the workloads have been characterized, the next steps are to specify each
traffic class's SLIs and SLOs (\Sec{background}) and describe the
network used by the simulator in \Sec{model}.
We use Dhall~\cite{dhall} as the configuration language because it allows for
typed and modular configurations, and it can be converted to widely
used formats like JSON and YAML.
Specifications can be fed to either the simulator itself through the standard
front end or to the weight optimizer.
If the simulator is invoked directly through the front end, it will simply run the
classes together according to the specification and return whether the SLOs are
met or violated, and why.
Alternatively, the optimizer will run the simulator many times in
search of a suitable weight allocation.

\subsection{Specification} \label{s:method-spec}

\begin{figure}[t]
    \lstset{
        numbers=left,
        numbersep=8pt,
        numberstyle=\tiny\color{gray},
        xleftmargin=0.5cm,
        xrightmargin=0.2cm,
    }
    \begin{lstlisting}[]
-- Top-level configuration
{ network =
  { link = T.Gbps 100
  , rtt = T.Us 10
  }
, queue = Queue.FIFO
, cc = ./dctcp.dhall
, classes = [ ./foo.dhall, ./bar.dhall ]
} : Config

-- In `dctcp.dhall`
CC.Normal
  { r_init = T.Gbps 100
  , u_target = 1.0
  , thresh = T.KB 100
  , beta = 1.0
  , eta = 5.5
  }

-- In `foo.dhall`
{ name = "Foo"
, flowsizes = Workload.Cdf "websearch.txt"
, interarrivals =
    Workload.Lognormal
      { mu = 11.3, sigma = 2.0 }
, slis =
  [ { name = "tail_slowdown"
    , metric = Metric.Percentile 0.99
    , attr = Attr.Slowdowns
    -- ... (snip)
    }
  ]
, slo = "tail_slowdown < 3.0"
}
-- `bar.dhall` is similar
  \end{lstlisting}
    \caption{\small \bf An abridged specification with two traffic classes.}
    \label{fig:spec}
\end{figure}

\Fig{spec} shows an example specification for a network configuration, traffic
classes, SLIs, and SLOs.
First, we provide the link capacity and the round-trip time (lines 3--4),
followed by the queueing discipline used at the bottleneck (line 6).
The congestion control protocol is imported from \texttt{dctcp.dhall}
(reproduced in lines 12--18). The meanings of the congestion control
parameters are described in more detail in the next section.
Likewise, the traffic classes \texttt{Foo} and \texttt{Bar}
are imported from \texttt{foo.dhall} and \texttt{bar.dhall}, respectively.

We define a traffic class \texttt{Foo} on line 21.
\texttt{Foo}'s message size distribution is given by the cumulative distribution in
\texttt{websearch.txt}, as described in Homa~\cite{homa}.
For interarrival times, we approximate them using a log-normal
distribution with mean $\mu$ equal to 11.3 and shape parameter $\sigma$ equal to
2.0. Together, the mean interarrival time and message size specify the
average bandwidth required for this traffic class; the interarrival and message
distribution, along with the congestion control protocol, control the burstiness
of traffic at the bottleneck link.

To evaluate this traffic class's performance, we select the 99th percentile
slowdown as our SLI.
We can also filter over flow size ranges in the SLI definition, but this is
omitted from this example for brevity.
The last step for \texttt{Foo} is to form an SLO by writing a logical predicate over
the SLI (line 33).
Here, we say that the P99 slowdown should be less than 3$\times$. We could also
combine different SLIs with different thresholds, e.g., so that small messages
can have a tighter bound than longer transfers.

The example includes a second traffic class \texttt{Bar}, defined in much the same
way, but with independent choices for traffic size distribution, interarrival burstiness,
SLIs, and SLOs.

Once the specification is written, the user can pass it to swp to ask it
to make SLO predictions under simulation.
The front end will:

\begin{enumerate}
    \item Read and validate the specification.
    \item Use the specification to build and run a simulation, which will
          terminate with a set of statistics.
    \item Digest the statistics to populate the SLIs.
    \item Predict whether or not each SLO will be met.
\end{enumerate}

\noindent A key contribution is to make simulations fast enough that a large space of
configurations and SLOs can be quickly explored.
Upon completion, swp will produce a set of predictions.
For each traffic class, it predicts a value for each SLI as well as a final
prediction for the SLO.
A slightly modified version of the specification can also be passed to the
weight optimizer, which we describe next.

\begin{algorithm}[t]
    \small
    \caption{Weight optimization loop \label{alg:optim}}
    \begin{algorithmic}[1]
        \Let{baselines}{\Call{FindBaselines}{classes}}
        \Let{weights}{\Call{Normalize}{baselines}} \Comment{class to weight}
        \Let{success}{\texttt{False}}
        \While{not timed out}
        \Let{losses}{\Call{GetLosses}{classes, weights}} \Comment{simulate}
        \Let{losses}{\Call{Sort}{losses}} \Comment{(class, loss) pairs sorted by loss}
        \If{all losses negative}
        \Let{success}{\texttt{True}}
        \State \textbf{break}
        \EndIf
        \Let{$\_$, $l_{\text{min}}$}{losses[0]}
        \If{$l_{\text{min}} > 0$}
        \State \textbf{break} \Comment{fail when min loss is positive}
        \EndIf
        \Let{rev}{\Call{reverse}{losses}}
        \Let{iterator}{\Call{zip}{losses, rev}} \Comment{iterate from both ends}
        \For{($k$, $l_{k}$), ($q$, $l_{q}$) \textbf{in} iterator}
        \If{$l_{k} \geq 0$ \textbf{or} $l_{q} \leq 0$}
        \State \textbf{break}
        \EndIf
        \Let{$\Delta$}{$|l_{k} / 2| \cdot \text{weights}[k]$} \Comment{transfer amount}
        \Let{weights[$k$]}{weights[$k$] - $\Delta$}
        \Let{weights[$q$]}{weights[$q$] + $\Delta$}
        \EndFor
        \EndWhile
        \State \textbf{return} weights, success
    \end{algorithmic}
\end{algorithm}

\subsection{A Weight Optimizer} \label{s:method-optim}

The final component of swp is an optimizer that hooks into the
simulator's Rust API and searches for switch weights that can meet the SLOs
of a set of traffic classes.
To start, each traffic class defines a loss function that takes the simulation
output and quantifies the distance between the observed SLI value and its
target SLO threshold.
For the purpose of tail latency SLOs, we assume the threshold is an upper
bound, so if the value exceeds the threshold, then the class requires more
capacity, but if the SLI is under the threshold, then the class potentially has
extra slack that can be given to other classes.
To compute loss, we simply use
\begin{equation}
    \text{loss} = -\frac{\text{SLO threshold} - \text{SLI value}}{\text{SLO threshold}}.
\end{equation}

\noindent Here, a negative loss indicates the SLO is met, and a positive loss
indicates the SLO is missed.

To find a starting point for the search process, we define
a \emph{baseline weight allocation} for each class:
the minimum normalized switch weight a class requires to meet its SLO assuming
worst case behavior by all other competing traffic classes---that is, that
all other traffic classes always have a packet queued at the bottleneck switch.
We use a binary search for this; each class's baselines are found in
parallel.

Once the baselines are found, we normalize them such that they sum to one, and
then we enter the optimization loop.
At each step of the loop, we first compute the losses by running the classes
together in simulation and then sorting their losses in ascending order.
Then in an inner loop we iterate through the losses from both ends
simultaneously, with the minimum loss class paired with the maximum loss class,
and so on.
If class $k$ is paired with class $q$, and if $k$ met its SLO while $q$ did
not, then we simply transfer weight from $k$ to $q$ in proportion to $k$'s
loss.
The optimization succeeds when all losses are negative, and it fails when
either 1) no loss is negative or 2) the optimization loop times out.
Pseudocode for the optimization loop is shown in \Alg{optim}.

%% file: model.tex
\section{A Network Model} \label{s:model}
Because it runs as the inner loop of swp,
our network model is designed to be as simple as possible while still
making accurate predictions about the aggregate tail SLO behavior seen in a network.
Like any model, it will necessarily diverge from reality. Our goal, however, is
not to represent the network stack with full fidelity but to isolate and
represent the aspects are most significant for whether tail SLOs are met by a particular
configuration.

The need for something that is simple and fast is motivated in part by our experience with
ns-3~\cite{ns-3}---a thorough and widely-used network simulator---whose
attention to detail lends it high fidelity, but whose explicit modeling of protocol
mechanisms prevents rapid answers to what-if questions.
ns-3 simulates every packet arrival and departure at every link, queue, and host, along with
host timeouts and the full host network protocol stack, with multiple events to send a packet
through transport, network, and link layers. Particularly for establishing confidence
intervals in tail behavior with bursty workloads, this level of detail can make simulations
take hours rather than minutes. Using ns-3 would also place a limit on how quickly swp could
react to measured changes in workload.

% For example, to configure a simulation of a datacenter network in ns-3, one
% must typically build a topology, configure error rates for links \ma{cut error rate -- it's almost irrelevant in datacenter simulations}, specify
% routes, select and set parameters for a congestion control algorithm, toggle
% numerous options for switches and RDMA, and explicitly assign IP addresses
% and port numbers. \ma{This configuration work is a burden on the person running the simulations but isn't the reason for long running times. That's because ns3 simulates every event (packets, timeouts) at every link, queue, host, etc. and it models a full host network stack (e.g., requiring multiple events to send a packet through transport, network, and link layers and out of an interface)}
%
% The level of detail results in a long running
% time, which can take up to several hours depending on the experiment,
% and which is further multiplied if one wants to run it enough times to
% establish a tight confidence interval.

We believe that a model of the network can be useful even if it ignores most of
the above details, especially if it is only used to reason about aggregate statistical behavior.
The central challenge, however, is defining one that adequately balances
simplicity against fidelity.
One of our contributions is a functional, rather than an operational, model
of the network to try to better strike this balance.

\subsection{Overview}

\begin{figure}[t]
    \centering
    \includegraphics[width=.75\linewidth]{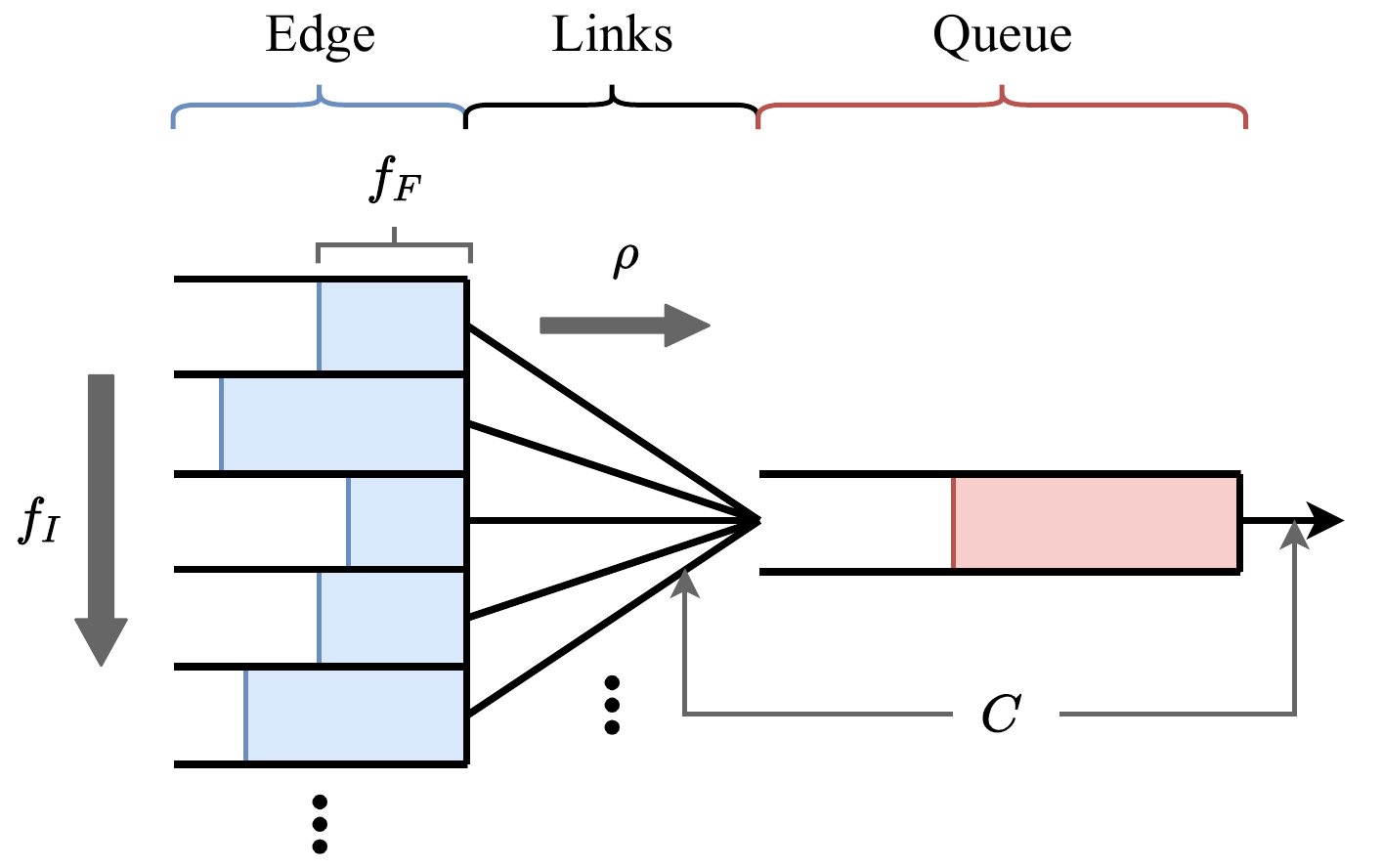}
    \vspace{-2mm}
    \caption{\small \bf
        A simple network model with a single bottleneck where all links have
        capacity $\boldsymbol{C}$. Flows arrive according to some flow size and interarrival
        time distribution, and each flow $\boldsymbol{i}$ sends at some rate $\boldsymbol{\rho_{i} \leq C}$.
    }
    \label{fig:swp-model}
    \vspace{-4mm}
\end{figure}

\Fig{swp-model} depicts the network model.
All links have capacity $C$, and there is a single bottleneck with an infinite
queue, following some queueing discipline.
Flows are generated at the edge with flow sizes and interarrival times drawn
from distributions $f_{F}$ and $f_{I}$.
Upon arrival, each flow $i$ begins sending data at some rate $\rho_{i}$
governed by a model of congestion control (described in the next section).
The delay between sources at the edge and the bottleneck is $\tau$ (giving a
round-trip time of $2\tau$).

Restricting the model to include only a single bottleneck queue in this manner
is a simplification, but meeting SLOs in even this simple case is nontrivial.
Moreover, where congestion is caused by incast or fan-in, there is only one
bottleneck, and Google has previously reported that last hop congestion is
the most common in datacenters~\cite{jupiter}.
In Section~\Sec{model-eval}, we compare predictions made by simulating against
this model to those made by ns-3 on a single bottleneck link.

\subsection{Congestion Control} \label{s:model-cc}

While usually complex, congestion control algorithms often have a functional
description that can be succinctly summarized.
For example, a particular algorithm will react to congestion signals, choose a
balance between utilization and queueing, and converge to a target
bandwidth allocation after some number of steps.
Our goal is to design a model that captures these characteristics.
Throughout, we use the notation $x^{+}$ to mean $\max(0, x)$.
First, for any given flow, its \emph{ideal} sending rate at time $t$ is denoted
$R(t)$, and it is given by

\begin{equation} \label{eq:Rt}
    R(t) =
    \begin{cases}
        r_{\text{init}} & t \leq 2\tau     \\
        r(t)^{+}        & \text{otherwise} \\
    \end{cases},
\end{equation}

\noindent where $r(t)$ is an update rule which will be described shortly.
Recall that $\tau$ is the one-way delay between the sources and the
bottleneck.
Upon arrival, a new flow begins sending immediately at some initial rate
$r_{\text{init}}$, and it will continue sending at this rate until it receives
congestion control feedback at time $2\tau$.
Many modern datacenter protocols, such as DCQCN~\cite{dcqcn} and HPCC~\cite{hpcc},
set $r_{\text{init}}$ to the line rate $C$. For protocols with a smaller $r_{\text{init}}$
value, we assume the initial window is paced~\cite{tcppacing} rather than ack clocked.
% \ma{maybe say here that traditionally transports have used a small $r_{init}$ (e.g., a few MSS / 2$\tau$), but some datacenter transports (e.g., DC-QCN, and probably HPCC - need to double check the paper) start at line rate}
%
We model the feedback delay explicitly because it has important implications
on the number of uncontrolled bytes in the network. As link speeds increase and
flow size distributions skew smaller, a greater fraction of traffic is transmitted
in an uncontrolled manner~\cite{bfc,aeolus}.

\paragraph{Update rule.}

The update rule $r(t)$ defines the core of the control loop, and it applies to
a flow as soon as the flow first receives feedback.
At this time, we say the flow is \emph{controlled}.
Likewise if a flow has begun sending but has not yet received feedback, we say
it is \emph{uncontrolled}.
Given this terminology, a high-level and idealized description of the update
rule is

\begin{equation*}
    \text{rate} =
    \frac{
        \text{total capacity} - \text{uncontrolled rate} - \text{queue drain}
    }{
        \# \text{ of controlled flows}
    },
\end{equation*}

\noindent where ``uncontrolled rate'' refers to the total sending rate across all
uncontrolled flows.
This rule follows our intuition that the rate at which a controlled flow should
send at any given time depends on how much \emph{residual} capacity there is at
the bottleneck---and that, in turn, depends on 1) how much uncontrolled traffic
there is and 2) how much queueing has accumulated at the switch.
For simplicity we do not model differences in fairness, so the residual
capacity is simply divided evenly among all controlled flows.

We build on this idealized version by introducing delays in signal propagation
as well as parameters for approximating different congestion control
algorithms.
We begin by defining the uncontrolled rate more precisely.
Let $i$ be an arbitrary flow, $S_{i}$ be its initial size, and $t_{\text{a}, i}$
be its arrival time.
The uncontrolled rate contributed by this flow $i$ at time $t$, called
$r_{\text{u}, i}(t)$, is given by

\begin{equation} \label{eq:rui}
    r_{\text{u}, i}(t) =
    \begin{cases}
        \dfrac{\min(r_{\text{init}} \cdot 2\tau, S_{i})}{2\tau} & 0 \leq t - t_{\text{a}, i} \leq 2\tau \\
        0                                                       & \text{otherwise}                      \\
    \end{cases}.
\end{equation}

\noindent In other words, each flow will contribute up to $r_{\text{init}} \cdot 2\tau$
uncontrolled bytes upon arrival, paced out over $2\tau$ time.
The total uncontrolled rate at time $t$ is then just the sum over the
individual flows,

\begin{equation}
    r_{\text{u}}(t) = \sum_{i} r_{\text{u}, i}(t).
\end{equation}

\noindent Next, the number of controlled flows at time $t$ is simply the number
of flows that have already sent $r_{\text{init}} \cdot 2\tau$ bytes and are
still sending at time $t$:

\begin{equation}
    N(t) = \big| \{i \mid t - t_{\text{a}} > 2\tau \text{ and } i \text{ has unsent bytes}\} \big|.
\end{equation}

\noindent And finally, let $Q(t)$ be the number of bytes queued at the
bottleneck at time $t$.

We are now ready to state the update rule $r(t)$.
Recalling that $C$ is the total link capacity, the update rule for each
controlled flow is

\begin{equation} \label{eq:rt}
    r(t) = \frac{
        U \cdot C - \beta \cdot r_{\text{u}}(t - 2\tau) - \dfrac{(Q(t - \tau) - T)^{+}}{2\tau}
    }{
        N(t - 2\tau)
    },
\end{equation}

\noindent where $U$, $\beta$, and $T$ are parameters to the congestion control
model.
Different settings will approximate different congestion control algorithms.

The parameter $U$ is the target utilization, and $\beta$ is always zero or one,
controlling whether or not the model reacts to the uncontrolled rate.
These parameters are meant to model an algorithm like HPCC~\cite{hpcc}, which
1) tries to keep near-zero queues by intentionally underutilizing links, and 2)
can detect congestion without waiting for queueing to occur.
On the other hand, algorithms like DCTCP~\cite{dctcp}, DCQCN~\cite{dcqcn}, and
TIMELY~\cite{timely} can only detect congestion after a queue builds up.
In this case, $\beta$ would be set to zero, and the threshold at which the
model reacts to queueing is controlled by $T$.

We also note the time delays on $r_{\text{u}}$, $Q$, and $N$.
These reflect our intuition about which signals are collected at the sources
($t - 2\tau$), and which signals are collected at the switch ($t - \tau$).
For example, a new flow that arrives cannot have its uncontrolled rate affect
the rates assigned to other flows until $2\tau$ after its arrival.
However, since queueing is measured at the bottleneck, the buildup of a queue
can be indicated to sources after $\tau$ time.

\paragraph{Convergence speed.}

Recall that $R(t)$ from \eqref{eq:Rt} is the ideal sending rate of a flow at
time $t$, as determined by the parameters and the model.
The \emph{actual} rate assigned to flows $\rho(t)$ will converge to $R(t)$ on a certain timescale dictated by the convergence speed of the congestion control algorithm. The convergence timescale is controlled by a smoothing parameter $\eta$:
\begin{equation}
    \frac{d\rho}{dt} = \frac{R(t) - \rho(t)}{\eta\tau}.
\end{equation}
This differential equation is the continuous-time equivalent of applying an exponentially weighted moving average (first-order low pass filter) to $R(t)$ to derive $\rho(t)$. Higher values of $\eta$ will result in slower convergence.
A full listing of the model's congestion control parameters are given in
\Tab{params}.

% \ma{One thing that's ambiguous here is when the smoothing starts. I assume its after the first $2\tau$, but do we do it starting from $r_{init}$ or the first rate computed by the congestion control algorithm using eq. (6) (i.e. after the source receives its first feedback). I think it makes more sense to start smoothing from the rate obtained after 1 RTT, rather than $r_{init}$.}
% \kz{Right now it starts at $r_{init}$; the idea was to go down gradually from the initial rate. Can fix though if need be, it probably won't change results much.} 
% \ma{probably fine to keep it as is. the other one just somehow seems more reasonable to me, since rinit is picked in a blind way anyway, so it doesn't matter that we deviate from it quickly. anyway, it actually might have some impact on the results for short flows (but longer than an RTT).}

\begin{table}
    \small
    \centering
    \begin{tabular}{p{2.5cm}p{3.6cm}}
        \toprule
        Model parameter   & Description                   \\
        \midrule
        $r_{\text{init}}$ & Initial send rate             \\
        $U$               & Target utilization            \\
        $T$               & Queue threshold               \\
        $\beta$           & Uncontrolled traffic reaction \\
        $\eta$            & Convergence smoothing         \\
        \bottomrule
    \end{tabular}
    \caption{\small \bf swp's congestion control parameters.}
    \label{tab:params}
    \vspace{-4mm}
\end{table}

\paragraph{Multiple traffic classes.}

The above model assumes there is only one class of traffic, but algorithms like
DCTCP can be adapted to the case where there are multiple traffic
classes, and each class is associated with a fixed scheduling weight at the switch.
Here, we describe how to generalize the model to cover this scenario.
To state our goal concretely, suppose there are $n$ classes, and suppose $k$ is
an arbitrary class---with weight $w_{k}$---that is continuously backlogged on
some time interval.
Then in the congestion control model, we want the capacity $C_{k}$ that is
available to class $k$ to be such that

\begin{equation}
    C_{k} \geq \frac{w_{k}}{\sum_{l=1}^{n} w_{l}} C
\end{equation}

\noindent on that time interval. Moreover, if a class is not using the link,
its capacity should be divided among all active classes in proportion to their
weights.

First, we restrict congestion signals $r_{\text{u}}$, $Q$, and $N$ to only
include information about a particular class $k$, yielding $r_{\text{u}, k}$,
$Q_{k}$, and $N_{k}$.
For example, $r_{\text{u}, k}$ would be the total rate of uncontrolled traffic
due to class $k$.
\footnote{
    Not to be confused with $r_{\text{u}, i}$ in \eqref{eq:rui}. For clarity,
    we have tried to use $i, j$ to index flows and $k, l$ to index classes.
}
\noindent Now we define the model's notion of an active class.
We say a class $k$ is \emph{active} at time $t$ when at least one of two
conditions holds:

\begin{enumerate}
    \item $Q_{k}(t) > 0$, or
    \item $k$ has a packet on the bottleneck wire at time $t$.
\end{enumerate}

% \tom{My argument here would be to define it in terms of an instantaneous virtual clock}
% \ma{The idea of a packet being on the wire at time $t$ is tricky to precisely define because till now this model has not depended on the simulator but checking condition 2 above relies on having a packet-level simulation. Not sure how best to explain this or if we should avoid getting into further details}

\noindent For convenience, we define a predicate $\text{active}_{k}(t)$ which
is true whenever the above criterion is met for class $k$.
We will also define a function $W$ which maps a set of class indices $K$ to the
sum of the class' weights,

\begin{equation}
    W(K) = \sum_{k \in K} w_{k}.
\end{equation}

\noindent With these definitions, we can write the bottleneck capacity
available to class $k$ at time $t$ as

\begin{equation}
    C_{k}(t) = \frac{w_{k}}{W(\{l \mid \text{active}_{l}(t)\} \cup \{k\})} C.
\end{equation}

\noindent Unlike in the original definition of $r(t)$ in \eqref{eq:rt}, we now
have a link capacity that is time varying.
Substituting this as well as the class-specific signals from above, we can
write the update rule for class $k$ as

\begin{equation} \label{eq:rkt}
    r_{k}(t) = \frac{
        U \cdot C_{k}(t - \tau) - \beta \cdot r_{\text{u}, k}(t - 2\tau) - \dfrac{(Q_{k}(t - \tau) - T)^{+}}{2\tau}
    }{
        N_{k}(t - 2\tau)
    }.
\end{equation}

\noindent Aside from the update rule, everything in the multi-class case works
in the same way as before.

% \ma{The discussion of multiple traffic classes could potentially go into an appendix with a short summary in the paper's body if we need the space.}

\subsection{Queueing Discipline}

We consider several standard queueing disciplines: first-in first-out (FIFO),
priority queueing (PQ), round robin (RR), deficit round robin (DRR), as well as
common hierarchical and weighted variants.
Some of these policies had long been considered impractical to implement on
switches, but recent work has shown how the mechanisms on newer programmable
switches can be used to approximate policies like weighted fair
queueing~\cite{cal-queues}.
We include a variety of policies in our model to observe their impact on SLOs.

% \subsection{Limitations} \todo{Write me}
\section{Evaluation} \label{s:model-eval}

\subsection{Evaluating the Network Model} \label{s:model-eval}

\begin{figure*}[t]
    \centering
    \begin{subfigure}[t]{0.24\linewidth}
        \centering
        \includegraphics[width=\textwidth]{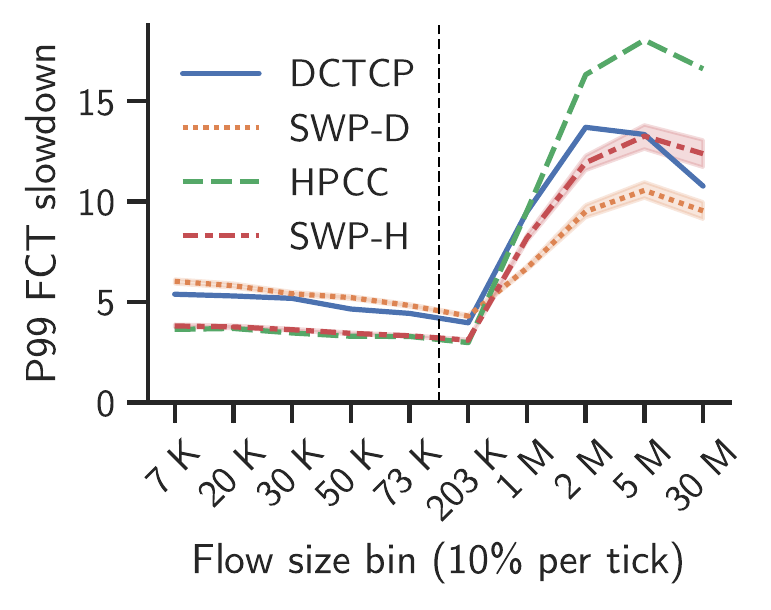}
        \caption{\bf WebSearch, 30\% utilization}
        \label{fig:ws-30}
    \end{subfigure}
    \hfill
    \begin{subfigure}[t]{0.24\linewidth}
        \centering
        \includegraphics[width=\textwidth]{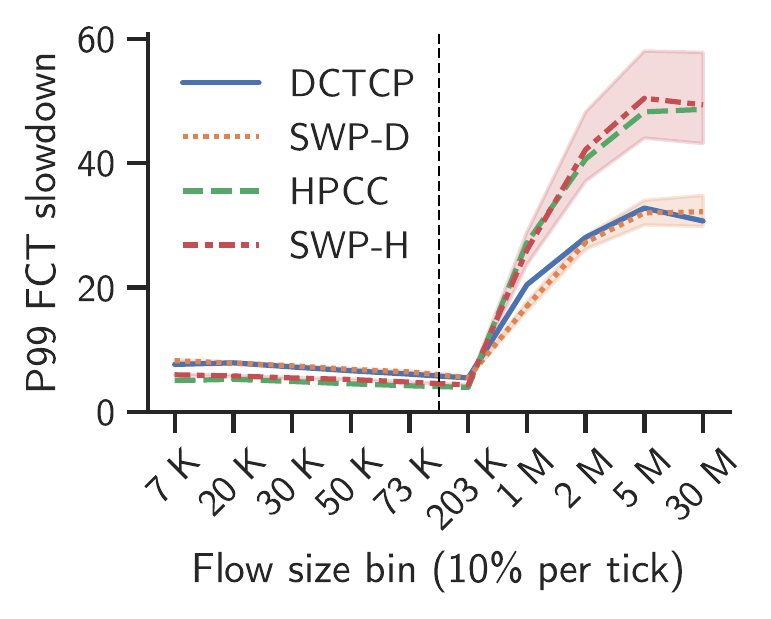}
        \caption{\bf WebSearch, 60\% utilization}
        \label{fig:ws-60}
    \end{subfigure}
    \hfill
    \begin{subfigure}[t]{0.24\linewidth}
        \centering
        \includegraphics[width=\textwidth]{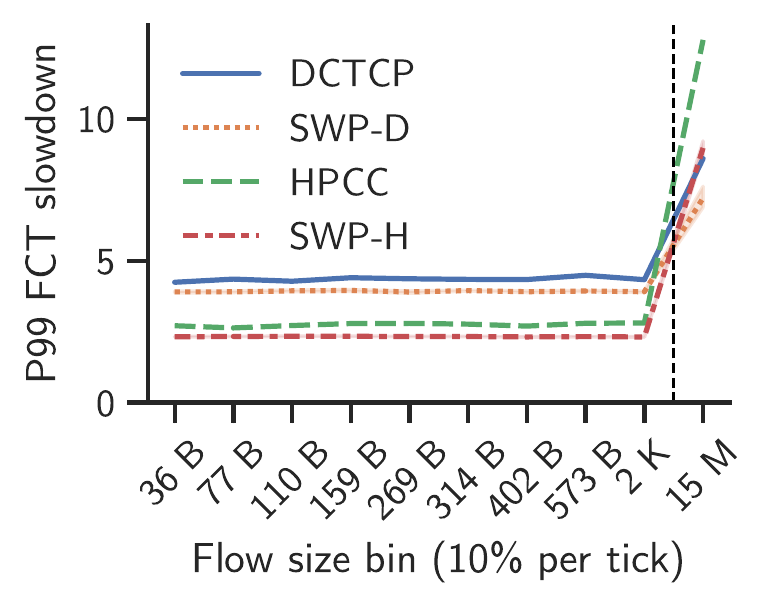}
        \caption{\bf Google, 30\% utilization}
        \label{fig:goog-30}
    \end{subfigure}
    \hfill
    \begin{subfigure}[t]{0.24\linewidth}
        \centering
        \includegraphics[width=\textwidth]{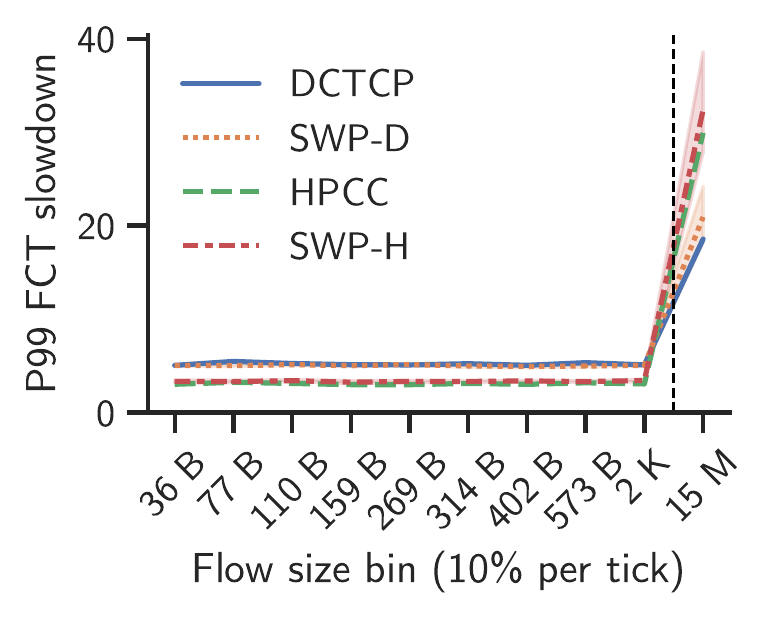}
        \caption{\bf Google, 60\% utilization}
        \label{fig:goog-60}
    \end{subfigure}
    \vspace{-2mm}
    \caption{\small \bf
        When there is a single bottleneck, swp's network model is effective at
        predicting the P99 slowdown of short flows under DCTCP and HPCC.
        Here, SWP-D and SWP-H are tuned to match the behavior of DCTCP and
        HPCC, respectively, by setting the congestion control parameters
        (\Tab{params}) to those shown in \Tab{validation-params}.
        Once the parameters are set, they are unchanged across load levels and
        flow size distributions.
        Binning is done according to flow size.
        In this experiment, there are ten equally-sized bins, and each tick on
        the horizontal axis is labeled with the largest flow size in the
        corresponding bin.
        The dashed vertical line approximately indicates one BDP.
    }
    \vspace{-2mm}
    \label{fig:validation}
\end{figure*}

\begin{table}
    \small
    \centering
    \begin{tabular}{p{2cm}p{1.5cm}p{1.5cm}}
        \toprule
        Parameter            & SWP-D           & SWP-H         \\
        \midrule
        $r_{\text{init}}$    & 100 Gbps        & 100 Gbps      \\
        $\boldsymbol{U}$     & \textbf{100\%}  & \textbf{90\%} \\
        $\boldsymbol{T}$     & \textbf{100 KB} & \textbf{0 KB} \\
        $\boldsymbol{\beta}$ & \textbf{0.0}    & \textbf{1.0}  \\
        $\boldsymbol{\eta}$  & \textbf{5.5}    & \textbf{5.0}  \\
        \bottomrule
    \end{tabular}
    \caption{\small \bf
        The congestion control parameters used for SWP-D and SWP-H in
        \Fig{validation}.
        Bolded rows indicate where parameters differ between the two configurations.
    }
    \label{tab:validation-params}
    \vspace{-4mm}
\end{table}

\begin{figure}
    \centering
    \includegraphics[width=0.8\linewidth]{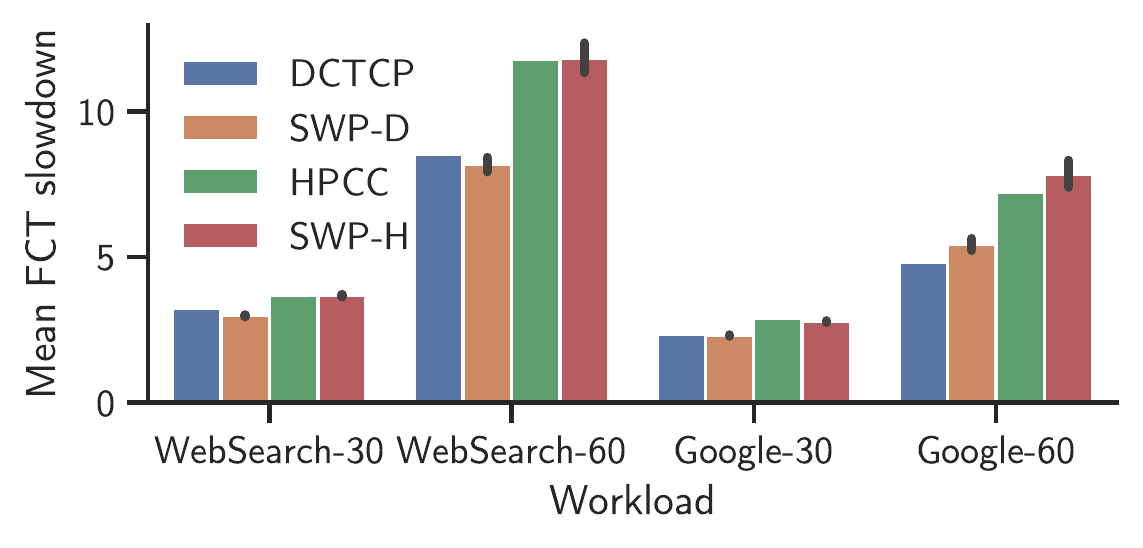}
    \vspace{-4mm}
    \caption{\small \bf
        For flows larger than a megabyte, SWP-D and SWP-H can approximate the
        mean slowdown given by DCTCP and HPCC.
        Labels on the horizontal axis give the flow size distribution and the
        utilization.
    }
    \label{fig:averages}
    \vspace{-4mm}
\end{figure}

\begin{table}
    \small
    \centering
    \begin{tabular}{l r r r}
        \toprule
        Workload     & ns-3   & swp  & Slowdown                            \\
        \midrule
        WebSearch-30 & 113.22 & 1.68 & \textbf{67.39}$\boldsymbol{\times}$ \\
        WebSearch-60 & 124.69 & 2.35 & \textbf{53.06}$\boldsymbol{\times}$ \\
        Google-30    & 54.77  & 0.67 & \textbf{81.75}$\boldsymbol{\times}$ \\
        Google-60    & 55.95  & 0.80 & \textbf{69.94}$\boldsymbol{\times}$ \\
        \bottomrule
    \end{tabular}
    \caption{\small \bf
        swp's simplified model allows a simulator to simulate 50,000 flows
        over 50$\boldsymbol{\times}$ faster than ns-3 in a comparable scenario.
        Here, ns-3 is running DCTCP and swp is running SWP-D.
        Running times are shown in minutes.
    }
    \label{tab:bench}
    \vspace{-4mm}
\end{table}

Before using swp to predict SLOs, we first evaluate our network model by comparing its predictions to the
output of ns-3 simulating a single bottleneck link.
For a higher degree of confidence, we try different flow size distributions at
different load levels using both DCTCP and HPCC, and we analyze tail latencies
across fine-grained flow size bins.
The goal of the model is not to report accurate predictions for per-flow
statistics, but rather to capture information about \emph{aggregate} statistics
like the averages and tails of flow completion times (FCTs).
This is what allows us to distill complex behavior into a simple model and
achieve quicker answers to the questions we are trying to ask.

Specifically, we evaluate the model along three axes:

\begin{enumerate}
    \item How well does the model predict the tail FCT slowdown of short flows?
    \item How well does the model predict the average FCT slowdown of long
          flows?
    \item How much faster is the model when
          compared to ns-3?
\end{enumerate}

\noindent For short flows, we believe the model should accurately predict tail slowdowns, since
those will depend largely on the aggregate behavior of the congestion control
model defined in \Sec{model-cc}.
However, since we explicitly do not model differences in fairness among long
flows, we do not always expect accurate tail predictions for them.
For these flows, we instead evaluate the model's ability to predict accurate
average slowdowns.
This reflects common practice, since short messages often require low,
predictable latency while long ones should achieve acceptable throughput.

In these experiments, we use a 100 Gbps bottleneck link and a 10
\si{\micro\second} round trip time (RTT).
We configure ns-3 with multiple sources sending to the same destination through
a bottleneck, similarly to the model shown in \Fig{swp-model}.
We use two publicly available flow size distributions: one is a web search
application~\cite{dctcp}, and the other is an aggregated workload from a Google
datacenter~\cite{homa}.
The workloads are run with a bursty log-normal interarrival time distribution
(shape parameter $\sigma = 2$) at two different load levels, 30\% and 60\%.
We also run them with both DCTCP and HPCC.

\Fig{validation} shows the tail FCT slowdown predictions produced by the model
across these configurations and across equally-sized flow size bins.
For short flows, the model is able to accurately predict the slowdowns at the
99th percentile, and at 60\% utilization it can even reasonably predict the
tail slowdown of long flows.
At low utilization, the tail predictions for long flows are less accurate
because 1) the model does not model short-term unfairness for long flows and 2)
at low utilization events have higher variance, and this increases the tail
variance of long flows.
%~\todo{Double check this explanation...}
%
The average FCT slowdown for long flows, however, remains accurate across load
levels.
This is shown in \Fig{averages}.
Lastly, the model arrives at these predictions up to 80$\times$ faster
than does ns-3.
\Tab{bench} shows the running time of ns-3 running DCTCP compared against
swp's model, with measurements taken on an Intel Xeon E5-2680 CPU.
The swp simulator simulates 50,000 flows up to 81$\times$ faster than does
ns-3.
Running times for HPCC and its corresponding swp model are similar.

%% file: eval.tex
\subsection{Evaluating Network Configurations}

% doing this \pg{Another (maybe easy) addition is to look at one of the configurations of \Fig{min-bw} and explain what weights you observed (for weights + FIFO), and how the min capacity is lower than sum of min capacities if the traffic classes were run in isolation. }

\begin{figure*}[t]
    \centering
    \begin{subfigure}[t]{0.24\linewidth}
        \centering
        \includegraphics[width=\textwidth]{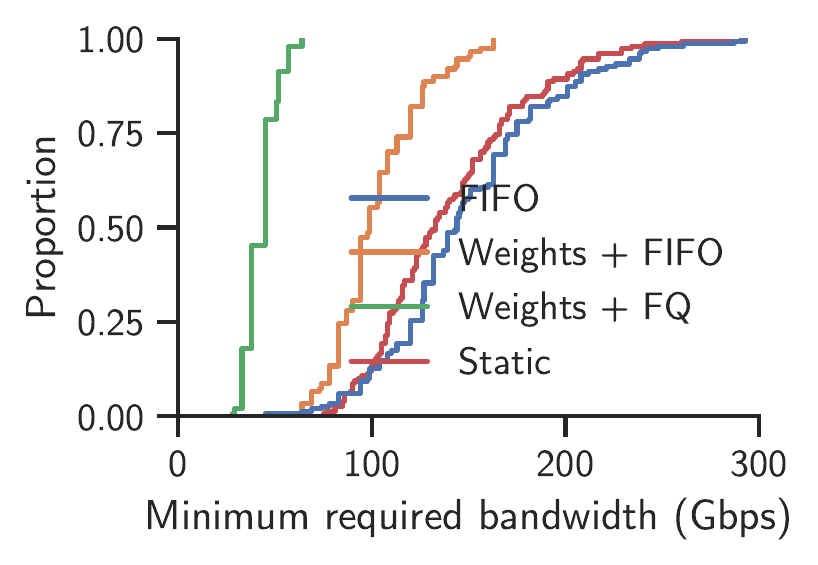}
        \caption{\bf All configurations}
        \label{fig:min-bw-3-all}
    \end{subfigure}
    \hfill
    \begin{subfigure}[t]{0.24\linewidth}
        \centering
        \includegraphics[width=\textwidth]{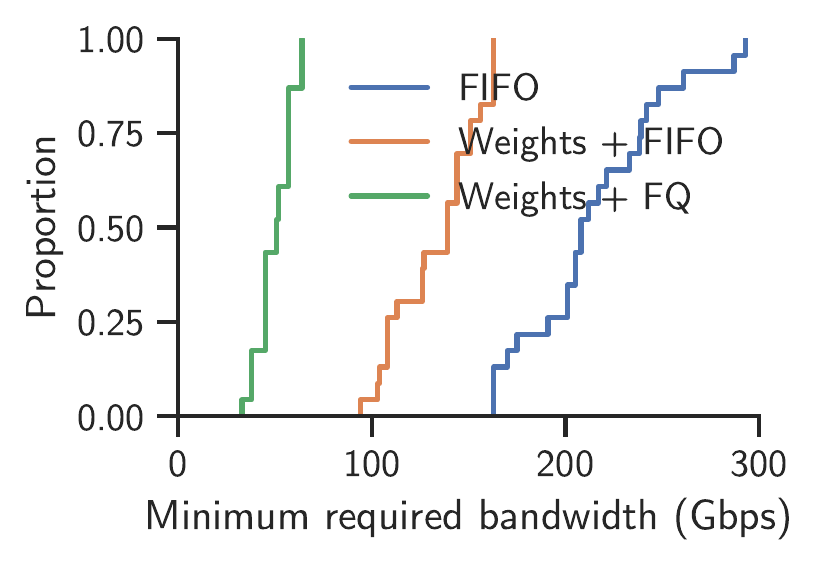}
        \caption{\bf
            Configurations with at least one tight SLO (threshold $<$ 4) on a
            bursty class ($\sigma>$ 1.7)}
        \label{fig:min-bw-3-tight-bursty}
    \end{subfigure}
    \hfill
    \begin{subfigure}[t]{0.24\linewidth}
        \centering
        \includegraphics[width=\textwidth]{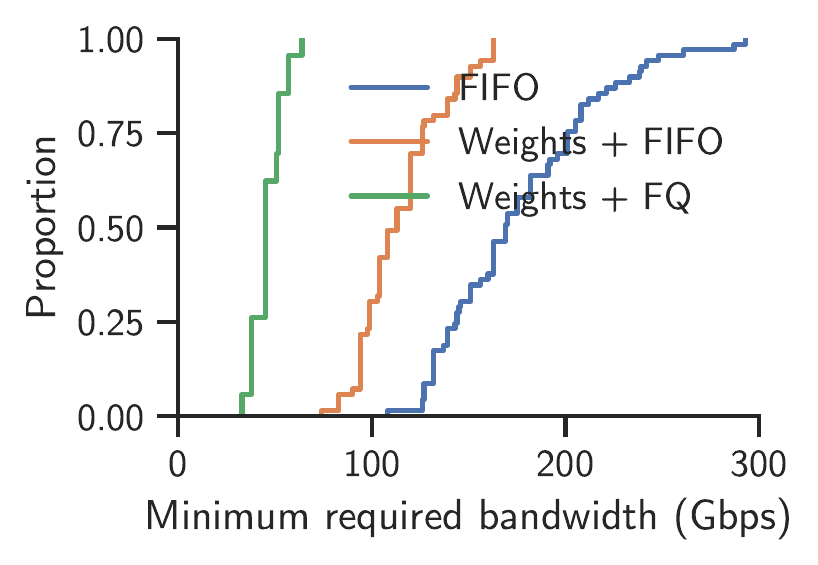}
        \caption{\bf
            Configurations with at least one tight SLO (threshold $<$ 4)
        }
        \label{fig:min-bw-3-tight}
    \end{subfigure}
    \hfill
    \begin{subfigure}[t]{0.24\linewidth}
        \centering
        \includegraphics[width=\textwidth]{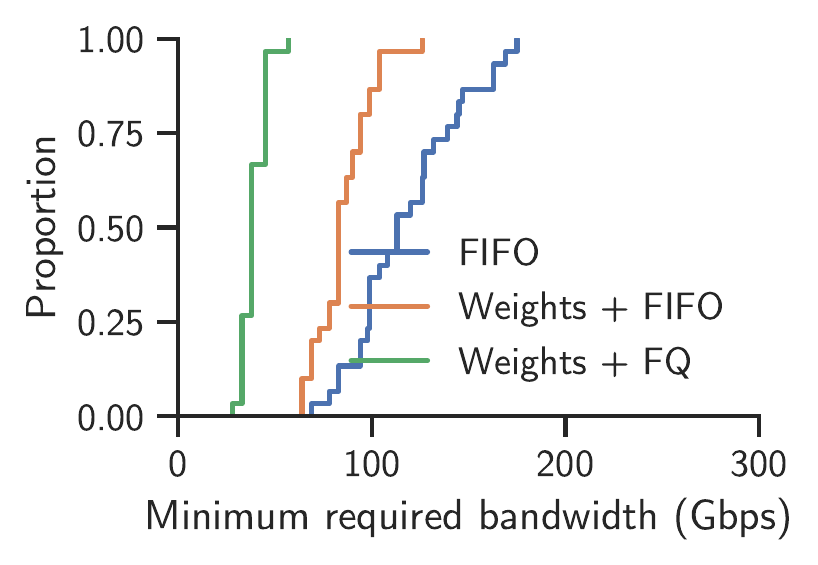}
        \caption{\small
            \bf Configurations with no tight SLOs and lower burstiness (all
            thresholds $\geq$ 4 and all $\boldsymbol{\sigma \leq 1.7}$)
        }
        \label{fig:min-bw-3-loose-smooth}
    \end{subfigure}
    \vspace{-2mm}
    \caption{\small \bf
        swp finds weights that can meet a set of SLOs using significantly
        less bandwidth than would be required with pure FIFO queueing.
        The advantage of weights over FIFO is yet more pronounced with tighter
        SLOs and higher application burstiness.
    }
    \label{fig:min-bw-3}
    \vspace{-2mm}
\end{figure*}

\begin{figure}[t]
    \centering
    \begin{subfigure}[t]{0.49\linewidth}
        \centering
        \includegraphics[width=\textwidth]{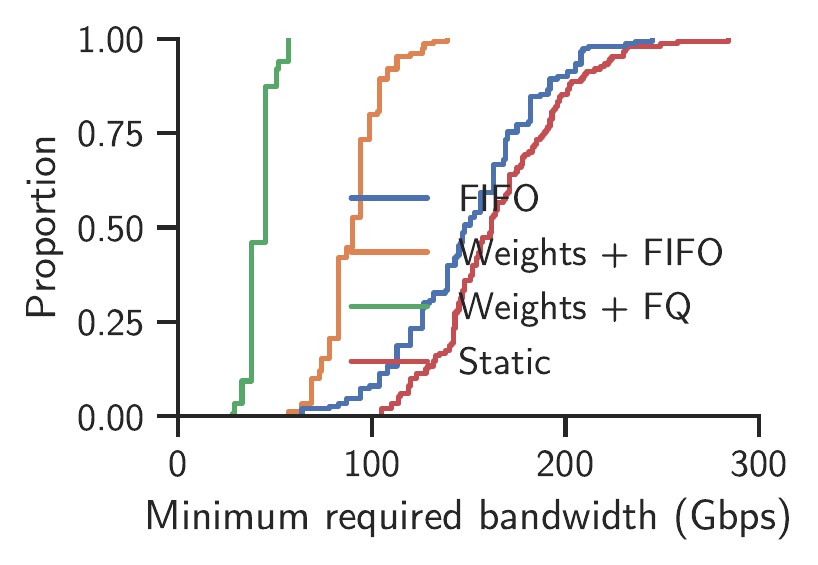}
        \caption{\bf All configurations}
        \label{fig:min-bw-5-all}
    \end{subfigure}
    \hfill
    \begin{subfigure}[t]{0.49\linewidth}
        \centering
        \includegraphics[width=\textwidth]{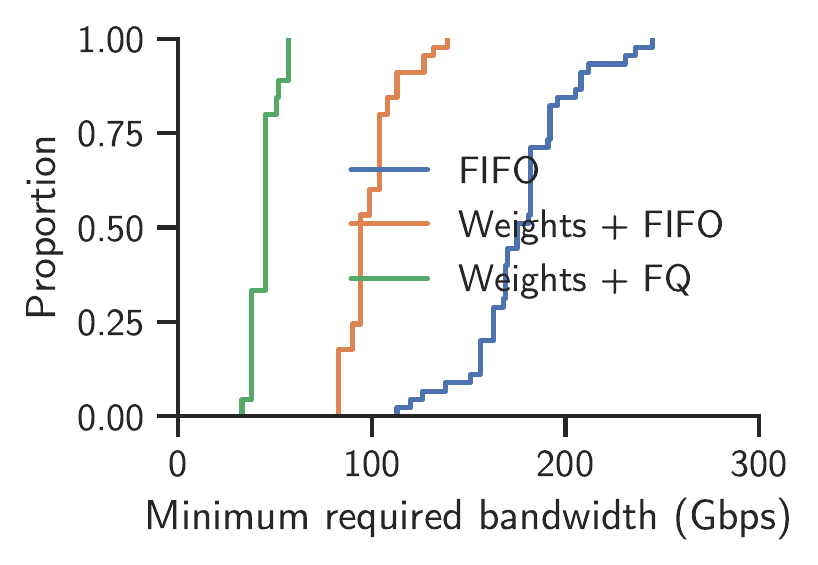}
        \caption{\bf
            Configurations with at least one tight SLO (threshold $<$ 4) on a
            bursty class ($\sigma>$ 1.7)}
        \label{fig:min-bw-5-tight-bursty}
    \end{subfigure}
    \caption{\small \bf
        When we choose five random traffic classes instead of three, the
        benefit of optimizing weights increases. Moreover, statically
        provisioning each class for the worst case becomes costlier than using
        pure FIFO.
    }
    \label{fig:min-bw-5}
\end{figure}

\begin{table}
    \small
    \centering
    \begin{tabular}{l l}
        \toprule
        Parameter             & Sample space              \\
        \midrule
        Flow sizes            & Google, Facebook, Alibaba \\
        Burstiness ($\sigma$) & [1.0, 2.0]                \\
        Mean rate (3-class)   & [5 Gbps, 10 Gbps]         \\
        Mean rate (5-class)   & [3 Gbps, 6 Gbps]          \\
        SLO threshold         & [3.0, 8.0]                \\
        \bottomrule
    \end{tabular}
    \caption{\small \bf
        The sample space for the random configurations used in \Fig{min-bw-3}
        and \Fig{min-bw-5}.
        The Google flow size distribution is the same as the one in
        \Sec{model-eval}, and the Facebook and Alibaba distributions are taken
        from the public repository of HPCC~\cite{hpcc-repo}.
        Burstiness is the shape parameter $\boldsymbol{\sigma}$ for the log-normal
        interarrival time distribution.
    }
    \label{tab:sample-space}
    \vspace{-5mm}
\end{table}

\begin{figure}
    \centering
    \includegraphics[width=0.9\linewidth]{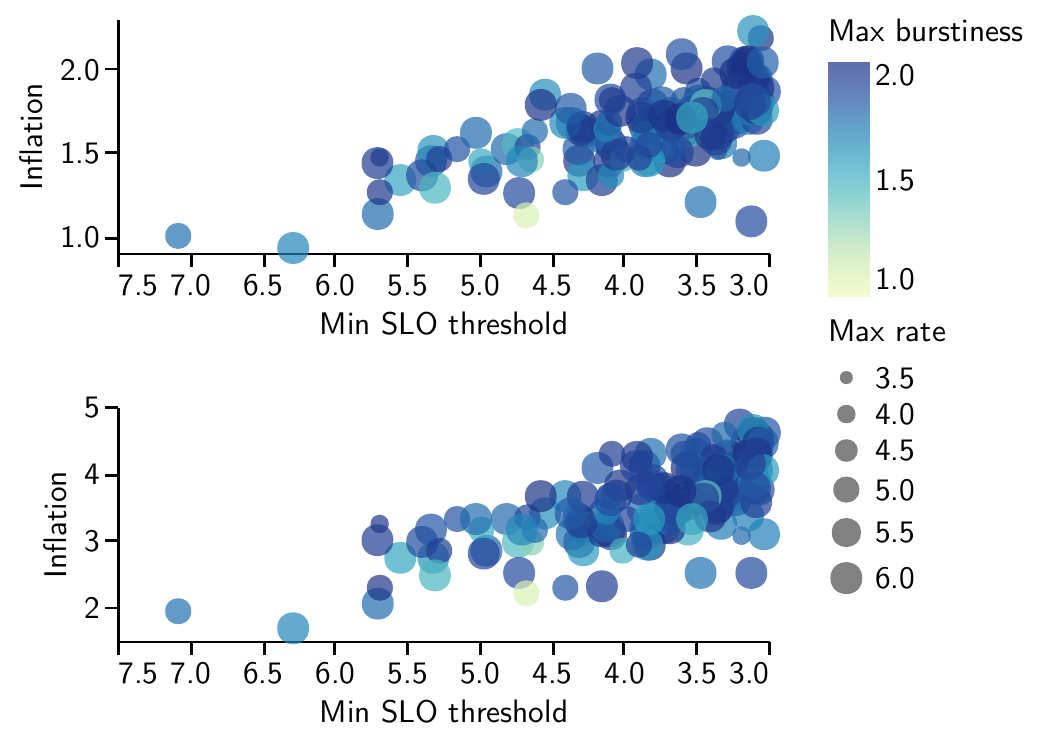}
    \caption{\small \bf
        Minimum bandwidth inflation of FIFO over Weights+FIFO (top)
        and Weights+FQ (bottom), where inflation is defined as the quotient
        of the minimum bandwidth requirements.
        Each circle is a configuration with five traffic classes.
        For each one, we track the minimum SLO threshold, the maximum
        burstiness, and the maximum mean rate.
        Here we see that the tighter the SLO threshold, the higher the inflation
        on average.
    }
    \label{fig:scatter}
    \vspace{-4mm}
\end{figure}

We next use the optimizer described in \Sec{method-optim} to
evaluate swp's ability to identify switch configurations that meet
target SLOs. Whether a particular configuration can be satisfied by
swp, a FIFO queue, or both is a coarse-grained metric.  Instead,
we generate a randomly generated set of scenarios, and consider
the minimum total bandwidth at the bottleneck that is sufficient to
meet the combined SLO for that configuration.
%
% Specifically given a set of traffic classes with their SLOs, swp should be
% able to find switch weights that will meet those SLOs with a lower link
% capacity than would be required with pure FIFO queueing.
%
Lower required bandwidth for the same scenario is better---it implies that the same
SLOs can be met with higher link utilization on a fixed bandwidth link.  We do not
consider priority scheduling in this experiment as that would require
extremely high link capacities to meet the target SLOs,
requiring very long runtimes to reach convergence at the tail.

We compare shared FIFO, per-class scheduling weights chosen by swp
with FIFO within each class, and the same with fair queueing within
each class. We consider scenarios
with three simultaneous traffic classes, and to cover a
wide range of cases, we sample traffic class characteristics randomly.
For each class, we uniformly select at random a flow size distribution from three
measured datacenter workloads, a
burstiness (log-normal $\sigma$) between 1 - 2, a mean sending rate between 5-10 Gbps,
and an SLO threshold for 99\% tail latency slowdown between 3 - 8$\times$.
We divide transfers into those that are smaller than the bandwidth-delay
product for this network (125KB) and those that are larger. The tail latency
slowdown is enforced against both sets independently, but with twice the slowdown
threshold for larger flows. This is to allow solutions that favor small flows, without
unduly starving medium-sized flows.
\Tab{sample-space} summarizes the sample space from which the random values are drawn.
We use a 10 \si{\micro\second} round trip time, and where congestion control
applies, we use the model of DCTCP.

For each of 150 randomly chosen scenarios, we use swp to search for the
minimum link capacity required to simultaneously meet all three SLOs for
both small and large flows, separately for per-class FIFO and an idealized
version of per-class fair queueing.
We also use binary search to identify the minimum link capacity needed to meet the SLOs
with a shared FIFO queue at the bottleneck.
%
% If weighting is used, we consider both FIFO and FQ for its internal queues
% (with congestion control turned off in the FQ case), and we use the optimizer
% to automatically find suitable weights to meet the SLOs.

\Fig{min-bw-3-all} shows the cumulative distribution function for
the minimum link capacity under different
strategies and traffic scenarios.
Averaged across all configurations,
using FIFO alone requires on average 44\% more link bandwidth
than if swp is used to optimize per-class FIFO weights, and an average of 247\%
more bandwidth than with swp with optimized hierarchical weighted fair queueing.

If we restrict the configurations to those with at least one bursty class
($\sigma > 1.7$) with a tight SLO (threshold $<$ 4), the average gap between
swp and FIFO widens to 62\% (\Fig{min-bw-3-tight-bursty}).
\Fig{min-bw-3-tight} shows the
results for configurations with SLO $<$ 4, but no restriction on burstiness,
while \Fig{min-bw-3-loose-smooth} shows the results with all SLOs $\geq$ 4 and
burstiness $\leq$ 1.7. The greatest advantage for swp comes on more challenging
scenarios.

% Note that for one of the 150 scenarios, FIFO does slightly better than swp.
% Recall that we assume congestion control is applied per-class. This means that
% in scenarios with low burstiness and loose SLOs, FIFO will sometimes outperform
% per-class weights.  With pure FIFO, a long flow in one class will sometimes
% back off due to queueing from competing classes, reducing the capacity needed
% to meet SLOs for those competing classes. With per-class weights, each class
% competes only against flows from its own class, and that sometimes harms
% performance.

We can also ask: how important is the swp optimizer described in Algorithm
\ref{alg:optim} to the benefits of our approach?
To study this, we consider a static version of swp, where we compute the bandwidth
needed for each traffic class to meet its SLOs independently, assuming no knowledge
of the behavior of the other traffic classes.  We then run this on the same scenarios
as we considered above. The result is plotted on the same graph
in \Fig{min-bw-3-all}, labeled as static.  The benefits of swp
relative to FIFO roughly disappear---that is, the advantage of per-class SLOs comes
from being able to take advantage of the cumulative slack across traffic classes.

Next, we consider configurations with five random traffic classes instead of
three, where we keep aggregate load the same by adjusting per-class mean rates to be
from 3 to 6 Gbps.
The result is shown in \Fig{min-bw-5}.
With more traffic classes, there is greater diversity of requirements,
presenting more opportunities for optimization.
In this setting, using FIFO queueing alone now requires 65\% more link
bandwidth than using optimized FIFO weights (\Fig{min-bw-5-all}), and in the
challenging scenario of a tight SLO on a bursty class, that number increases to
79\% (\Fig{min-bw-5-tight-bursty}).
Statically provisioning bandwidth for the worst case is also now more costly
than pure FIFO queueing.

Finally, the inflation factor of a configuration is the ratio between the
bandwidth required to meet SLOs with FIFO, divided by (respectively) that
required by swp weights with per-class FIFO and swp with per-class fair
queueing. \Fig{scatter} provides a scatter plot of inflation factors for all
150 five-class configurations, with each configuration indexed by the value of
the tightest SLO, the greatest burstiness, and the greatest mean rate across
the five traffic classes.  In general, configurations with tighter SLOs show
greater benefit for carefully optimized per-class traffic weights.

%% file: related.tex
\section{Related Work}
Achieving quality of service in packet switched networks
is a long-standing and rich research area. Quality of service
is easiest to achieve when we can assume that switches or routers
can keep per-flow state, such as Autonet-2~\cite{an2}, Intserv~\cite{intserv},
and RSVP~\cite{rsvp}. Parekh and Gallager showed that packet latency bounds
could be provided given fair queueing in the network~\cite{fq,parekh-gallager}.
Others have shown it is possible to emulate
fair queueing without per-flow state at every router~\cite{csfq}. However,
because of the difficulty of implementing these approaches
at high speeds, most work in the Internet gravitated to the use of priorities~\cite{diffserv}
% and admission control~\cite{admissioncontrol}
to deliver quality of service. The underlying assumption was that only
a small amount of near constant-bit rate traffic would need prioritization.
By contrast, our work focuses on providing quality of service for datacenter
traffic, where most or almost all network requests have timeliness constraints
and the traffic demand is inherently bursty.

%\paragraph{Per-packet latency guarantees.}
\smallskip
\noindent
\textbf{Per-packet latency guarantees.}
The closest to our work are QJump~\cite{qjump} and Silo~\cite{silo}
as they both target providing quality of service guarantees in the datacenter.
Both use a leaky bucket at sources to ensure senders do not exceed
some average rate and burstiness allowance.
Then, QJump uses switch priorities to bound the network packet delays for latency
sensitive applications.
SILO on the other hand uses network calculus to develop a VM placement
algorithm that can ensure network queueing delays do not exceed some bound.
Like earlier work,
these approaches target individual packet delays, rather than application-level
metrics, and traffic shaping can impose significant delays at the source particularly
for bursty traffic.
Moreover, for large scale datacenter networks, tight worst-case bounds can only
be provided for a small fraction of the traffic.
Our goal with swp is to provide quality of service for arbitrary size messages
for most or all of the network traffic,
using probabilistic rather than deterministic guarantees.
% Instead, user requirements are often on flows or messages, and they are stated
% in terms of tail latency percentiles---these are the probabilistic guarantees
% that swp targets.

% Existing solutions: QJump, Silo doesn't extend beyond per-packet SLAs, to per-flow guarantees

\smallskip
\noindent
\textbf{Scheduling flows for low latency.}
% D3, pfabric, pdq (brighten) - schedule flows based on app criteria to meet a deadline
% d2tcp - take the deadline and use it to adjust the aggressiveness as it nears the deadline
A number of systems, such as D$^3$~\cite{d3}, D$^2$TCP~\cite{d2tcp}, and PDQ~\cite{pdq} use deadlines to schedule flows for low latency. pFabric~\cite{pfabric}
uses shortest remaining time first to reduce average latency for short flows.
A significant limitation with these approaches is that they require applications
to provide the size of each flow, before it starts, to be able to assign it a deadline
or scheduling priority. Many applications in wide use lack this information, and
in fact the UNIX socket API does not allow applications to specify it even if known.
In addition, these schemes require switch hardware modifications.
swp helps operators achieve SLOs without new switch hardware or changing applications,
provided aggregate information is available about the distribution of flow sizes
and burstiness.

Homa~\cite{homa} is a transport protocol that uses switch priorities and receiver-driven use of priority queues to achieve low tail latency for short messages with existing
hardware. Like these other approaches, however, Homa assumes the size of the flow
is available. It also lacks a mechanism for predicting what tail latency guarantees it
can provide, nor does it provide an algorithm for balancing SLOs across traffic classes
or flow sizes.
% While it is interesting to speculate whether Homa could be extended to provide similar
% properties to swp, that is beyond the scope of this paper.
%
% In Homa, short messages are always transmitted blindly, which is too
% inflexible to support SLOs across different flow size ranges.
%
% One goal of swp is to allow for such fine-grained SLOs, and to allow
% each traffic class to specify its own goals.
% homa - flow scheduler, order flows dynamically (only full bisection, only bottlenecks at the edge)
% (didn't study deadline based mechanisms)
% (requires flow size info)
% (prioritizes initial window, and that might not meet SLAs)

% leaving out for now
% \paragraph{Stochastic network calculus.}
% One inspiration for our work was a series of papers that examined
% the use of stochastic network calculus to provide probabilistic
% guarantees for tail latency in storage systems~\cite{sncmeister}.

%
% slo coflow scheduler (schedules a collection of flows)
% https://arxiv.org/pdf/1709.01384.pdf

%% file: conclusion.tex
\section{Conclusion}
In this paper, we have built and evaluated swp, a new tool for
quickly identifying network switch configurations
to achieve tight tail latency bounds for bursty
datacenter traffic patterns. Given measured data about
the distribution of message sizes and message interarrival times,
swp finds traffic-class based switch scheduling weights to
accomplish class-specific operator-defined service level objectives (SLOs).
A key innovation is a 50-80$\times$ faster network simulation engine that elides
detail but produces accurate estimates of tail behavior through
a bottleneck link for two popular
data center congestion control protocols, DCTCP and HPCC,
for a variety of switch scheduling algorithms.
We use swp on randomly chosen scenarios to show that swp
can identify switch configurations that meet target SLOs
at much lower bandwidth than FIFO.

    {\em This work does not raise any ethical issues.}